\documentstyle[epsfig, subfigure, lscape]{mn}

\def\ltapprox{\raise 2pt \hbox {$<$} \kern-1.1em \lower 5pt \hbox {$\approx$}}
\def\ltsim{\raise 2pt \hbox {$<$} \kern-1.1em \lower 4pt \hbox {$\sim$}}
\def\gtsim{\raise 2pt \hbox {$>$} \kern-1.1em \lower 4pt \hbox {$\sim$}}

\title{On the role of injection in kinetic approaches to nonlinear
particle acceleration at non-relativistic shock waves}

\author[Pasquale Blasi, Stefano Gabici \& Giulia Vannoni]
{P. Blasi$^1$, S. Gabici$^2$ and G. Vannoni$^2$\\
$^1$INAF/Osservatorio Astrofisico di Arcetri, Largo E. Fermi, 5 
I-50125 Firenze (Italy)\\
$^2$Max-Planck-Institut fuer Kernphysik, Saupfercheckweg, 1 D - 
69117 Heidelberg (Germany)}

\begin{document}
\maketitle

\begin{abstract}
The dynamical reaction of the particles accelerated at a shock front
by the first order Fermi process can be determined within kinetic models 
that account for both the hydrodynamics of the shocked fluid and the transport 
of the accelerated particles. These models predict the appearance of multiple 
solutions, all physically allowed. We discuss here the role of injection in 
selecting the {\it real} solution, in the framework of a simple phenomenological 
recipe, which is a variation of what is sometimes referred to as 
{\it thermal leakage}. 
In this context we show that multiple solutions basically disappear and when
they are present they are limited to rather peculiar values of the parameters.  
We also provide a quantitative calculation of the efficiency of particle 
acceleration at cosmic ray modified shocks and we identify the fraction of
energy which is advected downstream and that of particles escaping the 
system from upstream infinity at the maximum momentum. The consequences of
efficient particle acceleration for shock heating are also discussed.
\end{abstract}


\section{Introduction}\label{sec:intro}

Diffusive particle acceleration at non-relativistic shock fronts is an extensively studied 
phenomenon. Detailed discussions of the current status of the investigations 
can be found in some excellent reviews (Drury 1983; Blandford \& Eichler 1987;
Berezhko \& Krimsky 1988; Jones \& Ellison 1991; Malkov \& Drury 2001).
While much is by now well understood, some issues are still subjects
of much debate, for the theoretical and phenomenological implications that 
they may have. One of the most important of these is the reaction of the 
accelerated particles on the shock: the violation of the 
{\it test particle approximation} occurs when the acceleration process 
becomes sufficiently efficient that the pressure of the accelerated
particles is comparable with the incoming gas kinetic pressure.
Both the spectrum of the particles and the structure of the shock
are changed by this phenomenon, which is therefore intrinsically 
nonlinear.

At present there are three viable approaches to determine the reaction 
of the particles upon the shock: one is based on the ever-improving numerical
simulations (Jones \& Ellison 1991; Bell 1987; Ellison, M\"{o}bius \& Paschmann
1990; Ellison, Baring \& Jones 1995, 1996; Kang \& Jones 1997; Kang, Jones \& 
Gieseler 2002; Kang \& Jones 2005) that allow one to achieve a self-consistent 
treatment of several effects. 

The second approach is based on the so-called two-fluid model, and treats cosmic 
rays as a relativistic second fluid. This class of models was proposed and discussed
in (Drury \& V\"{o}lk 1980, 1981; Drury, Axford \& Summers 1982; Axford, Leer 
\& McKenzie  1982; Duffy, Drury \& V\"{o}lk 1994). These models allow one
to obtain the thermodynamics of the modified shocks, but do not provide
information about the spectrum of accelerated particles.
 
The third approach is semi-analytical and may be very helpful to understand 
the physics of the nonlinear effects in a way that sometimes is difficult
to achieve through simulations, due to their intrinsic complexity and
limitations in including very different spatial scales.
Blandford (1980) proposed a perturbative approach in which 
the pressure of accelerated particles was treated as a small perturbation.
By construction this method provides the description of the reaction 
only for weakly modified shocks.

Alternative approaches were proposed by Eichler (1984), Ellison \& Eichler
(1984), Eichler (1985) and Ellison \& Eichler (1985), based on the assumption 
that the diffusion of the particles is sufficiently energy dependent that 
different parts of the fluid are affected by particles with different energies. 
The way the calculations are carried out implies a sort of separate solution 
of the transport equation for subrelativistic and relativistic particles, so 
that the two spectra must be somehow connected at $p\sim mc$ {\it a posteriori}. 

In (Berezhko, Yelshin \& Ksenofontov 1994; Berezhko, Ksenofontov 
\& Yelshin 1995; Berezhko 1996) the effects of the non-linear reaction 
of accelerated particles on the maximum achievable energy were investigated, 
together with the effects of geometry. The maximum energy of the particles 
accelerated in supernova remnants in the presence of large acceleration 
efficiencies was also studied by Ptuskin \& Zirakashvili (2003a,b).

The need for a {\it practical} solution of the acceleration problem in the 
non-linear regime was recognized by Berezhko \& Ellison (1999), where a simple
analytical broken-power-law approximation of the non-linear spectra was presented. 

Recently, some promising analytical solutions of the problem of non-linear 
shock acceleration have appeared in the literature (Malkov 1997; Malkov, Diamond 
\& V\"{o}lk 2000; Blasi 2002, 2004). Blasi (2004) considered for the first time
the important effect of seed pre-existing particles in the acceleration region 
(the linear theory of this phenomenon was first studied by Bell (1978)). In a
recent work by Kang \& Jones (2005) the seed particles were included in 
numerical simulations of the acceleration process.

Numerical simulations have been instrumental to identify the dramatic
effects of the particles reaction: they showed that even when the fraction 
of particles injected from the thermal gas is relatively small, the energy 
channelled into these few particles can be an appreciable part of the kinetic 
energy of the unshocked fluid, making the test particle approach unsuitable. 
The most visible effects of the reaction of the accelerated particles on 
the shock appear in the spectrum of the accelerated particles, which shows
a peculiar hardening at the highest energies. The analytical approaches
reproduce well the basic features arising from nonlinear effects in
shock acceleration. 

There is an important point which is still lacking in the calculations of 
the non-linear particle acceleration at shock waves, namely the possible 
amplification of the background magnetic field, found in the numerical 
simulations by Lucek \& Bell (2000, 2000a) and Bell \& Lucek (2001) and 
recently described by Bell (2004). This effect is still ignored in all 
calculations of the reaction of cosmic rays on the shock structure. 
We will not include this effect in the present paper. 

Nonlinear effects in shock acceleration of thermal particles result in 
the appearance of multiple solutions in certain regions of the parameter
space. This phenomenon is very general and was found in both the two-fluid
models (Drury \& V\"{o}lk 1980, 1981) and in the kinetic models (Malkov
1997; Malkov et al. 2001; Blasi 2004). Monte Carlo approaches do not 
show multiple solutions. 

This behaviour resembles that of critical systems, with a bifurcation
occurring when some threshold is reached in a given order parameter. In 
the case of shock acceleration, it is not easy to find a way of discriminating 
among the multiple solutions when they appear. In (Mond \& Drury 1998), a two 
fluid approach has been used to demonstrate that when three solutions appear, 
the one with intermediate efficiency for particle acceleration is unstable to 
corrugations in the shock structure and emission of acoustic waves. Plausibility 
arguments may be put forward to justify that the system made of the shock plus the 
accelerated particles may sit at the critical point (see for instance the 
paper by Malkov, Diamond \& V\"{o}lk (2000)), but we are not 
aware of any real proof that this is what happens. The physical parameters 
that play a role in this approach 
to criticality are the maximum momentum achievable by the particles in the 
acceleration process, the Mach number of the shock, and the injection 
efficiency, namely the fraction of thermal particles crossing the shock that 
are accelerated to nonthermal energies.
The last of them, the injection efficiency, hides a crucial
physics problem by itself, and plays an important role in establishing 
the level of shock modification. This efficiency parameter in reality is 
determined by the microphysics of the shock and should not be a free parameter 
of the problem. Unfortunately, our poor knowledge of such microphysics, in 
particular for collisionless shocks, does not allow us to establish a
clear and universal connection between the injection efficiency and the
macroscopic shock properties. Put aside the possibility to have a fully
self-consistent picture of this phenomenon, one can try to achieve a 
phenomenological description of it. Kang, Jones \& Gieseler (2002)
introduced a sort of weight function to determine a return probability 
of particles in the downstream fluid to the upstream fluid, as a function
of particle momentum. Only sufficiently suprathermal particles can 
jump back to the upstream region and therefore take part in the acceleration 
process. Here we adopt an injection recipe which is similar to the
{\it thermal leakage} model of Kang et al. (2002) (see also previous papers by 
Malkov (1998) and by Gieseler et al. (2000)) and implement it in the
semi-analytical approach of Blasi (2002, 2004). We investigate then the 
phenomenon of multiple solutions and show that the injection recipe 
dramatically reduces the appearance of these situations. We also study
in some detail the efficiency for particle acceleration as a function 
of the Mach number of the shock and the maximum momentum of the 
accelerated particles. 

The paper is structured as follows: in section \ref{sec:nonlin} we 
briefly describe the method proposed by Blasi (2002) for the
calculation of the spectum and pressure of particles accelerated at
a modified shock. We describe the appearance of multiple solutions in
section \ref{sec:multiplesolutions}, and the comparison with the method of
Malkov (1997) in section \ref{sec:comparisonmalkov}. In section 
\ref{sec:injectionrecipe} we introduce a recipe for the injection of particles 
from the thermal pool. This recipe is then used in section 
\ref{sec:singlesolutions} 
to show how the regions of parameter space where multiple solutions appear are 
drastically reduced by the self-regulated injection. In section \ref{sec:escape}
we discuss the efficiency of particle acceleration at modified shocks,
and stress the role of escape of particles from upstream infinity. 
The consequences of the cosmic ray modification on the shock heating 
are investigated in section \ref{sec:heating}. We conclude in section
\ref{sec:conclusions}.

\section{A kinetic model for particle acceleration at modified shocks}
\label{sec:nonlin}

In this section we describe the method proposed by Blasi (2002, 2004) 
for the calculation of the spectrum and pressure of the particles
accelerated at a shock surface, when the reaction of the particles
is taken into account. No seed particles are included here.

The equation that describes the diffusive transport of particles in one 
dimension is
$$
\frac{\partial}{\partial x}
\left[ D  \frac{\partial}{\partial x} f(x,p) \right] - 
u  \frac{\partial f (x,p)}{\partial x} + 
$$
\begin{equation}
~~~~~~~~~~~~
\frac{1}{3} \frac{d u}{d x}~p~\frac{\partial f(x,p)}{\partial p} + Q(x,p) = 0,
\label{eq:trans}
\end{equation}
where we assumed stationarity ($\partial f/\partial t=0$). The $x$ axis is
oriented from upstream to downstream, as in fig. 1. The pressure of the 
accelerated particles slows down the fluid upstream before it crosses the 
shock surface, therefore the gas velocity at upstream infinity, $u_0$, is 
different from $u_1$, the fluid speed just upstream of the shock.
The injection term is taken in the form $Q(x,p)=Q_0(p) \delta(x)$.    
\begin{figure}
\resizebox{\hsize}{!}{\includegraphics{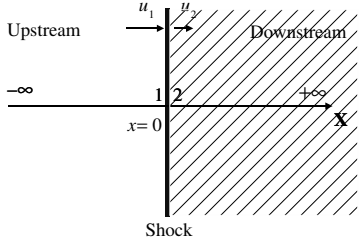}}
\caption[]{Schematic view of the shock region.} 
\label{fig:protons}
\end{figure}

As a first step, we integrate Eq. \ref{eq:trans} around $x=0$, from $x=0^-$ to
$x=0^+$, denoted in fig. 1 as points ``1'' and ``2'' respectively, so that the
following equation can be written:
\begin{equation}
\left[ D \frac{\partial f}{\partial x}\right]_2 -
\left[ D \frac{\partial f}{\partial x}\right]_1 +
\frac{1}{3} p \frac{d f_0}{d p} (u_2 - u_1) + Q_0(p)= 0,
\end{equation}
where $u_1$ ($u_2$) is the fluid speed immediately upstream (downstream) 
of the shock and $f_0$ is the particle distribution function at the shock
location.
By requiring that the distribution function downstream is independent of
the spatial coordinate (homogeneity), we obtain
$\left[ D \frac{\partial f}{\partial x}\right]_2=0$, so that the boundary 
condition at the shock can be rewritten as
\begin{equation}
\left[ D \frac{\partial f}{\partial x}\right]_1 =
\frac{1}{3} p \frac{d f_0}{d p} (u_2 - u_1) + Q_0(p).
\label{eq:boundaryshock}
\end{equation}
We can now perform the integration of Eq. \ref{eq:trans} from $x=-\infty$ to
$x=0^-$ (point ``1''), in order to take into account the boundary condition at 
upstream infinity. Using Eq. \ref{eq:boundaryshock} we obtain
$$
\frac{1}{3} p \frac{d f_0}{d p} (u_2 - u_1) - u_1 f_0 + Q_0(p)+
\int_{-\infty}^{0^-} dx f \frac{d u}{d x} + 
$$
\begin{equation}
~~~~~\frac{1}{3}\int_{-\infty}^{0^-} 
dx  \frac{d u}{d x} p \frac{\partial f}{\partial p} = 0.
\label{eq:step}
\end{equation}
We introduce the quantity $u_p$ defined as
\begin{equation}
u_p = u_1 - \frac{1}{f_0} \int_{-\infty}^{0^-} dx \frac{d u}{d x} f(x,p),
\label{eq:up}
\end{equation}
whose physical meaning is instrumental to understand the nonlinear 
reaction of particles. The function $u_p$ is the average fluid 
velocity experienced by particles with momentum $p$ while diffusing 
upstream away from the shock surface. In other words, the effect of the 
average is that, instead of a constant speed $u_1$ upstream, a particle 
with momentum $p$ experiences a spatially variable speed, due to the 
pressure of the accelerated particles. Since 
the diffusion coefficient is in general $p$-dependent, particles with 
different energies {\it feel} a different compression coefficient, higher 
at higher energies if, as expected, the diffusion coefficient is an 
increasing function of momentum (see (Blasi 2004) for further details
on the meaning of the quantity $u_p$). 

With the introduction of $u_p$, Eq. \ref{eq:step} becomes 
\begin{equation}
\frac{1}{3} p \frac{d f_0}{d p} (u_2 - u_p) - f_0 \left[u_p+\frac{1}{3} 
p \frac{du_p}{dp} \right] + Q_0(p) = 0 ,
\label{eq:step1}
\end{equation}
where we used the fact that
$$
p\frac{d}{dp}\int_{-\infty}^{0^-} dx \frac{du}{dx} f = 
p \left[\frac{df_0}{dp} (u_1-u_p) - f_0 \frac{du_p}{dp} \right].
$$
The solution of Eq. \ref{eq:step1} for a monochromatic injection at momentum
$p_{inj}$ is
$$
f_0 (p) =
$$
$$
\int_{p_0}^{p} \frac{d{\bar p}}{{\bar p}} 
\frac{3 Q_0({\bar p})}{u_{\bar p} - u_2} \exp\left\{-\int_{\bar p}^p 
\frac{dp'}{p'} \frac{3}{u_{p'} - u_2}\left[u_{p'}+\frac{1}{3}p' 
\frac{du_{p'}}{d p'}\right]\right\}=
$$
$$
\frac{3 R_{sub}}{R_{sub}-1} \frac{\eta n_{gas,1}}{4\pi p_{inj}^3} \times
$$
\begin{equation}
\times
\exp\left\{-\int_{p_{inj}}^p 
\frac{dp'}{p'} \frac{3}{u_{p'} - u_2}\left[u_{p'}+\frac{1}{3}p' 
\frac{du_{p'}}{d p'}\right]\right\}.
\label{eq:inje}
\end{equation}
Here we used $Q_0(p) = \frac{\eta n_{gas,1} u_1}{4\pi p_{inj}^2} 
\delta(p-p_{inj})$, with $n_{gas,1}$ the gas density immediately 
upstream ($x=0^-$) and $\eta$ the fraction of the particles crossing 
the shock which take part in the acceleration process.

Here we introduced the two quantities $R_{sub}=u_1/u_2$ and $R_{tot}=u_0/u_2$, 
which are respectively the compression factor at the gas subshock and 
the total compression factor between upstream infinity and downstream. 
For a modified shock, $R_{tot}$ can attain values much larger than
$R_{sub}$ and more in general, much larger than $4$, which is the maximum value
achievable for an ordinary strong non-relativistic shock. The increase of the 
total compression factor compared with the prediction for an ordinary shock is 
responsible for the peculiar flattening of the spectra of accelerated 
particles that represents a feature of nonlinear effects in shock acceleration.
In terms of $R_{sub}$ and $R_{tot}$ the density immediately upstream is 
$n_{gas,1}=(\rho_0/m_p)R_{tot}/R_{sub}$.

In Eq. \ref{eq:inje} we can introduce a dimensionless quantity 
$U(p)=u_p/u_0$ so that
$$
f_0^{inj} (p) = \left(\frac{3 R_{sub}}{R_{tot} U(p) - 1}\right) 
\frac{\eta n_{gas,1}}{4\pi p_{inj}^3} \times
$$
\begin{equation}
\times \exp \left\{-\int_{p_{inj}}^p 
\frac{dp'}{p'} \frac{3R_{tot}U(p')}{R_{tot} U(p') - 1}\right\}.
\label{eq:inje1}
\end{equation}

The nonlinearity of the problem reflects in the fact that $U(p)$ 
is in turn a function of $f_0$ as it is clear from the definition of $u_p$. 
In order to solve the problem we need to write the equations for the 
thermodynamics of the system including the gas, the cosmic rays 
accelerated from the thermal pool and the shock itself. 

The velocity, density and thermodynamic properties of the fluid
can be determined by the mass and momentum conservation equations, with the 
inclusion of the pressure of the accelerated particles. We write these 
equations between a point far upstream 
($x=-\infty$), where the fluid velocity is $u_0$ and the density is 
$\rho_0=m n_{gas,0}$, and the point where the fluid velocity is 
$u_p$ (density $\rho_p$).
The index $p$ denotes quantities measured at the point where the
fluid velocity is $u_p$, namely at the point $x_p$ that can be reached
only by particles with momentum $\geq p$ (this is clearly an approximation,
but as shown in section \ref{sec:comparisonmalkov} it provides a good 
agreement with other calculations where this approximation is not used). 

The mass conservation implies:
\begin{equation}
\rho_0 u_0 = \rho_p u_p.
\label{eq:mass}
\end{equation}
Conservation of momentum reads:
\begin{equation}
\rho_0 u_0^2 + P_{g,0} = \rho_p u_p^2 + P_{g,p} + P_{CR,p},
\label{eq:pressure}
\end{equation}
where $P_{g,p}$ is the gas pressure at the point $x=x_p$ and $P_{CR,p}$
is the pressure of accelerated particles at the same point (we use the 
symbol $CR$ to mean {\it cosmic rays}, in the sense of {\it accelerated 
particles}). The mass and momentum escaping fluxes in the form of accelerated particles
have reasonably been neglected. Note that at this point the equation for 
energy conservation has not been used. 

Our basic assumption, similar to that used in (Eichler 1984), is that
the diffusion is $p$-dependent and more specifically that the diffusion
coefficient $D(p)$ is an increasing function of $p$. Therefore the typical
distance that a particle with momentum $p$ travels away from the shock is
approximately $\Delta x\sim D(p)/u_p$, larger for high energy particles
than for lower energy particles\footnote{For the cases of interest, $D(p)$
increases with $p$ faster than $u_p$ does, therefore $\Delta x$ is a
monotonically increasing function of $p$.}. As a consequence, at each 
given point $x_p$ only particles with momentum larger than $p$ are able 
to affect appreciably the fluid. Strictly speaking the validity of the 
assumption depends on how strongly the diffusion coefficient depends on 
the momentum $p$ (see section \ref{sec:comparisonmalkov}).

Since only particles with momentum $\geq p$ can reach the point 
$x=x_p$, we can write
\begin{equation}
P_{CR,p} \simeq \frac{4\pi}{3} \int_{p}^{p_{max}} dp p^3 v(p) f_0(p),
\label{eq:CR}
\end{equation}
where $v(p)$ is the velocity of particles with momentum $p$, $p_{max}$ 
is the maximum momentum achievable in the specific situation under 
investigation.

From Eq. \ref{eq:pressure} we can see that there is a maximum distance, 
corresponding to the propagation of particles with momentum $p_{max}$ 
such that at larger distances the fluid is unaffected by the accelerated 
particles and $u_p=u_0$.

The equation for momentum conservation is:
\begin{equation}
\frac{dU}{dp} \left[ 1 - \frac{1}{M_0^2} U^{-(\gamma_g+1)} \right] + 
\frac{1}{\rho_0 u_0^2} \frac{d P_{CR}}{dp} = 0.
\end{equation}
Using the definition of $P_{CR}$ and multiplying by $p$, this equation 
becomes
\begin{equation}
p\frac{dU}{dp} \left[ 1 - \frac{1}{M_0^2} U^{-(\gamma_g+1)} \right] = 
\frac{4\pi}{3 \rho_0 u_0^2} p^4 v(p) f_0(p),
\label{eq:eqtosolve}
\end{equation}
where $f_0$ is known once $U(p)$ is known. 
Eq. \ref{eq:eqtosolve} is therefore an integral-differential nonlinear 
equation for $U(p)$. The solution of this equation also provides the 
spectrum of the accelerated particles.
 
The last missing piece is the connection between $R_{sub}$ and $R_{tot}$, the
two compression factors appearing in Eq. \ref{eq:inje1}. The compression 
factor at the gas shock around $x=0$ can be written in terms of the Mach 
number $M_1$ of the gas immediately upstream through the well known expression
\begin{equation}
R_{sub} = \frac{(\gamma_g+1)M_1^2}{(\gamma_g-1)M_1^2 + 2}.
\end{equation}
On the other hand, if the upstream gas evolution is adiabatic, then the Mach
number at $x=0^-$ can be written in terms of the Mach number of the fluid at
upstream infinity $M_0$ as
$$
M_1^2 = M_0^2 \left( \frac{u_1}{u_0} \right)^{\gamma_g+1} =
M_0^2 \left( \frac{R_{sub}}{R_{tot}} \right)^{\gamma_g+1},
$$
so that from the expression for $R_{sub}$ we obtain
\begin{equation}
R_{tot} = M_0^{\frac{2}{\gamma_g+1}} \left[ 
\frac{(\gamma_g+1)R_{sub}^{\gamma_g} - (\gamma_g-1)R_{sub}^{\gamma_g+1}}{2}
\right]^{\frac{1}{\gamma_g+1}}.
\label{eq:Rsub_Rtot}
\end{equation}

Now that an expression between $R_{sub}$ and $R_{tot}$ has been found, Eq.
\ref{eq:eqtosolve} basically is an equation for $R_{sub}$, with the boundary 
condition that $U(p_{max})=1$. Finding the value of $R_{sub}$ (and the 
corresponding value for $R_{tot}$) such that $U(p_{max})=1$ also provides 
the whole function $U(p)$ and, through Eq. \ref{eq:inje1}, the distribution 
function $f_0(p)$. If the reaction of the accelerated particles is small, 
the {\it test particle} solution is recovered. 

\section{The appearance of multiple solutions}
\label{sec:multiplesolutions}

In the problem described in the previous section there are several 
independent parameters. While the Mach number of the shock and the 
maximum momentum of the particles are fixed by the physical conditions
in the environment, the injection momentum and the acceleration efficiency 
are free parameters. The procedure to be followed to determine the solution 
was defined in (Blasi 2002): the basic problem is to find the value of 
$R_{sub}$ (and therefore of $R_{tot}$) for which $U(p_{max})=1$. In Fig. 2 
we plot $U(p_{max})$ as a function of $R_{tot}$, 
for $T_0=10^5 K$, $p_{max}=10^5 mc$ and $p_{inj}=10^{-2} mc$ 
in the left panel and $p_{inj}=10^{-3} mc$ in the right panel 
($m$ here is the mass of protons). The parameter 
$\eta$ was taken $10^{-4}$ in the left panel and $10^{-3}$ in the right panel.
The different curves refer to different choices of the Mach number
at upstream infinity. The physical solutions are those corresponding 
to the intersection points with the horizontal line $U(p_{max})=1$, so that 
multiple solutions occur for those values of the parameters for 
which there is more than one intersection with $U(p_{max})=1$. These 
solutions are all physically acceptable, as far as the conservation 
of mass, momentum and energy are concerned. 

\begin{figure*}
\label{fig:sol}
 \begin{center}
  \mbox{\epsfig{file=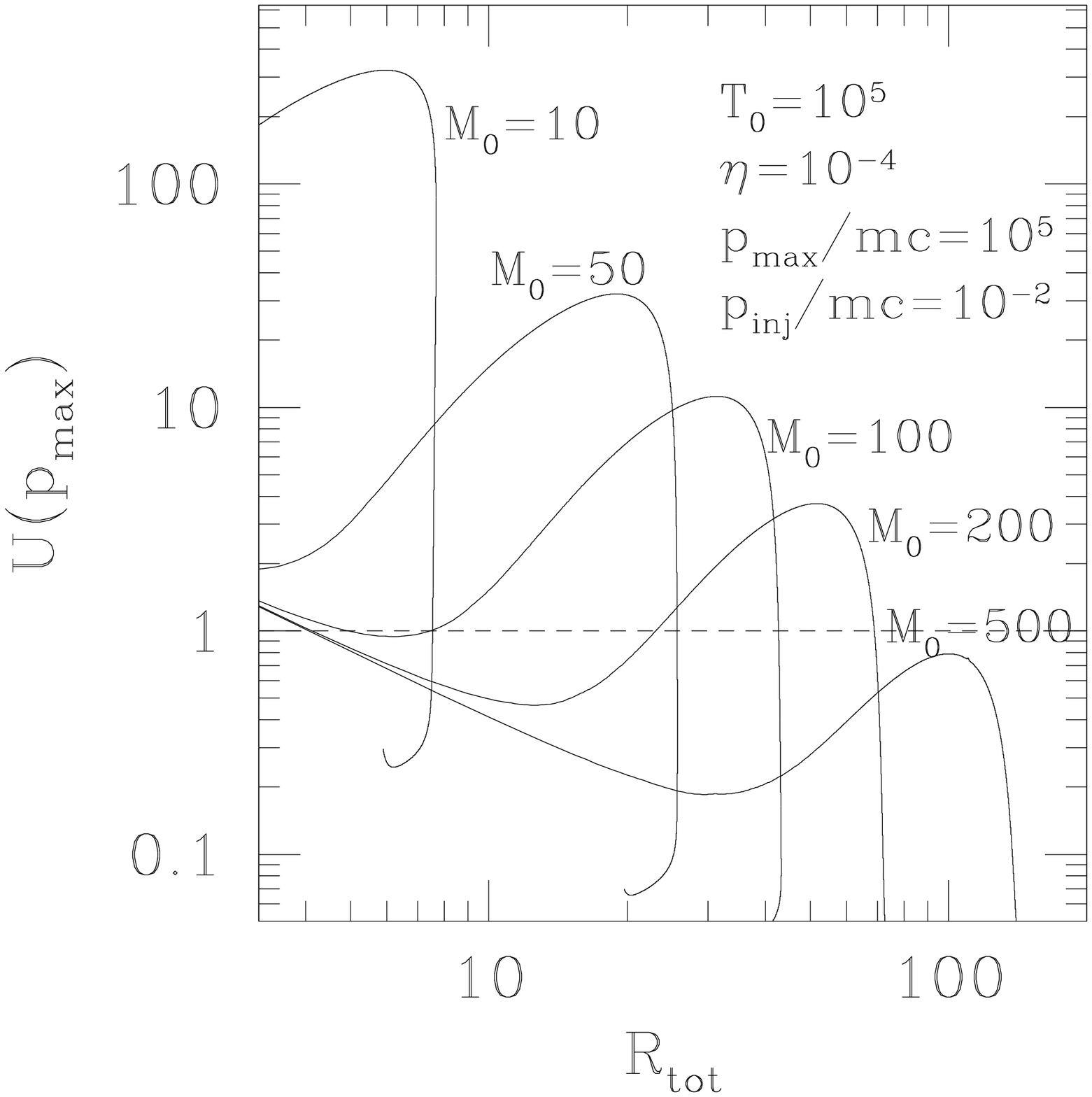 ,width=6.cm}}
  \mbox{\epsfig{file=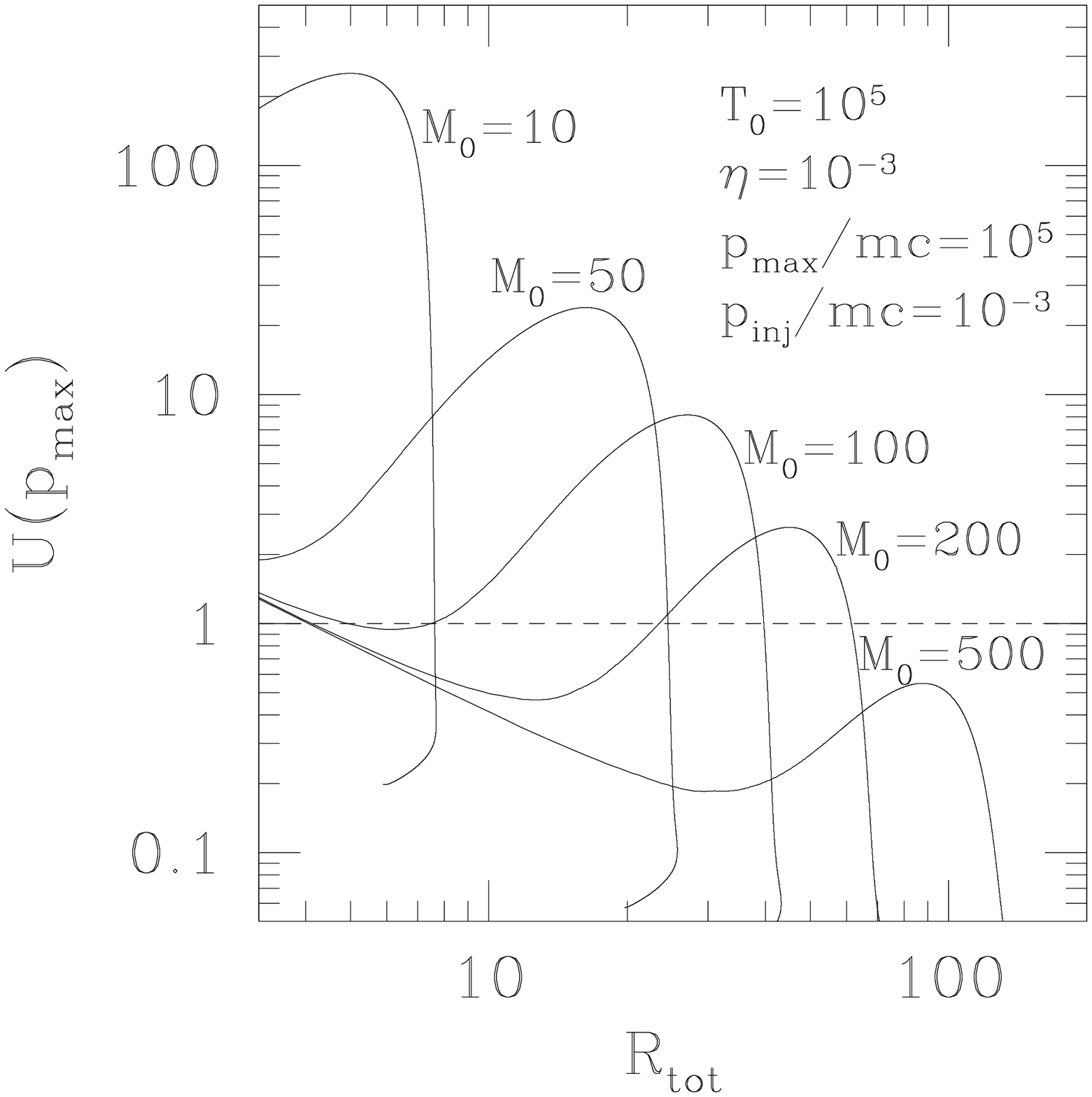 ,width=6.cm}}
  \caption{{\it Left panel}: $U(p_{max})$ as a function of the total compression 
factor for $T_0=10^5$ K, $\eta=10^{-4}$, $p_{max}=10^5~mc$ and 
$p_{inj}=10^{-2}~mc$ for the Mach numbers indicated. {\it Right panel}: same as in 
left panel but for $\eta=10^{-3}$ and $p_{inj}=10^{-3}~mc$. }
 \end{center}
\end{figure*}

It can be seen from both panels in Fig. 2 that for
low values of the Mach number, only one solution is found. This solution
may be significantly far from the quasi-linear solution. Indeed, for 
$M_0=10$ the solution corresponds to $R_{tot}\sim 8$, instead of the
usual $R_{tot}\sim 4$ solution expected in the linear regime.
Lower values of the Mach number are required to fully recover the 
linear solution. 

When the Mach number is increased, there is a threshold 
value for which three solutions appear, one of which is the 
quasi-linear solution. For very large values of the Mach number the 
solution becomes one again, and it coincides with the
quasi-linear solution.

\begin{figure*}
\label{fig:multi2}
 \begin{center}
  \mbox{\epsfig{file=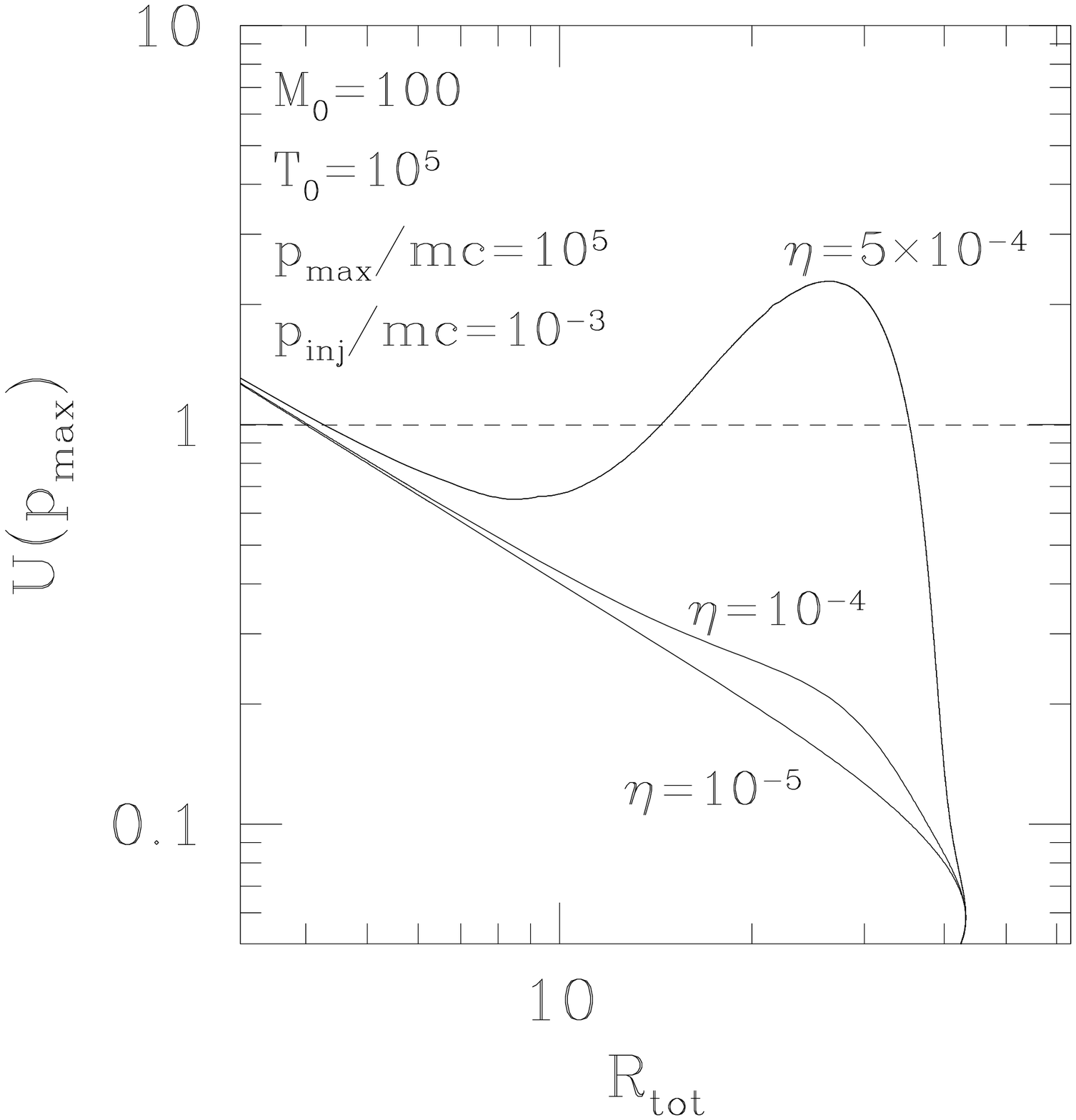 ,width=6.cm}}
  \mbox{\epsfig{file=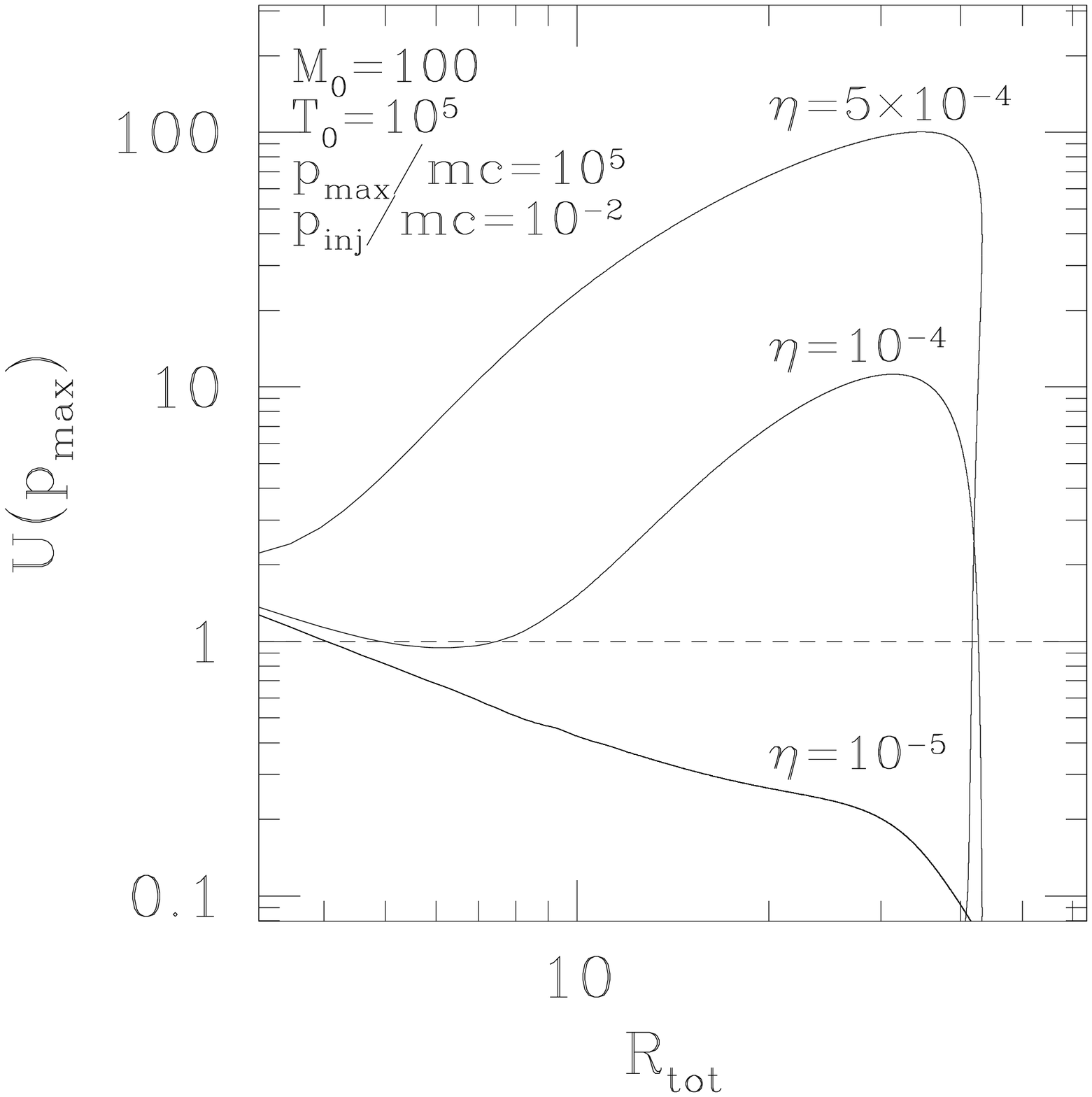,width=6.cm}}
  \caption{{\it Left panel}:$U(p_{max})$ as a function of the total compression 
factor for $T_0=10^5$ K, $p_{max}=10^5~mc$ and $p_{inj}=10^{-3}~mc$ at fixed 
Mach number $M_0=100$ for the efficiencies indicated. {\it Right panel}: same
as in left panel but for $p_{inj}=10^{-2}~mc$.}
 \end{center}
\end{figure*}

In Fig. 3 we show the appearance of the multiple 
solutions for the case $T_0=10^5 K$, $p_{max}=10^5 mc$ and 
$p_{inj}=10^{-3} mc$ with Mach number $M_0=100$ ($p_{inj}=10^{-2} mc$ and
$M_0=100$) in the left (right) panel. The curves here are obtained by 
changing the value of $\eta$. The same comments we made for Fig. 2 
apply here as well: low values of $\eta$ correspond
to weakly modified shocks, while for increasingly larger efficiencies
multiple solutions appear. The solution becomes one again in the limit
of large efficiencies, and it always corresponds to strongly modified
shocks.

The problem of multiple solutions is not peculiar of the kinetic
approaches to the non-linear theories of particle acceleration at
shock waves. The same phenomenon was in fact found initially in 
two-fluid models (Drury \& V\"{o}lk 1980, 1981), where however no 
information on the spectrum of the accelerated particles and on the 
injection efficiency was available.

\section{Comparison with an alternative approach}
\label{sec:comparisonmalkov}

Multiple solutions were also found by Malkov (1997) and Malkov et al. (2000), 
in the context of a semi-analytical kinetic approach. Aside from the 
technical differencies between that method and the one proposed 
by Blasi (2002, 2004), the main difference is in the fact that 
the former requires the knowledge of the exact expression for the 
diffusion coefficient as a function of the momentum of the particles, 
while the latter only requires that such diffusion coefficient is an 
increasing function of the particle momentum. While the first approach may 
provide us with an {\it exact} \footnote{Note however that even the approach of 
Malkov (1997) is based on several approximations: the solution is 
expanded to the first order and the contributions from gas pressure 
are ignored.} solution to the problem, the second is in 
fact more practical, in the sense of providing an approximate solution even
in those cases, the majority, in which no detailed information on the 
diffusion properties of the fluid is available. The solution provided in 
(Blasi 2002, 2004) is particularly accurate when the diffusion 
coefficient is Bohm-like, $D(p)=(1/3) r_L c$, expected in the case 
of saturated self-generation of waves in the vicinity of the shock
surface (Lagage \& Cesarsky 1983).

We will now discuss the quantitative comparison between the results 
of Malkov (1997) and those of Blasi (2002, 2004), by 
considering a single situation in which multiple solutions are 
predicted (in both approaches), and determining the spectra and 
compression factors in both methods. We start with briefly summarizing the 
approach of Malkov (1997). The following flow potential is introduced there:
\begin{equation}
\Psi = \int_x^0 dx^\prime u(x^\prime),
\end{equation}
which is used as a new independent spatial variable to replace $x$. Using the 
flow potential, it is possible to define an integral transformation of the flow 
profile as follows:
\begin{equation}
\label{inttra}
\hat{U}(p) = \frac{1}{u_0} 
\int^{-\infty}_{0^+} 
\exp \left[-\frac{q(p)}{3 D(p)} \Psi \right]
\frac{du}{dx} dx,
\end{equation}
where $q(p) = -d \ln f_0/d \ln p$ is the spectral index of the particle distribution 
function and $D(p)$ is the diffusion coefficient, which is assumed to be independent 
of the position. An integral equation for $\hat{U}(p)$ can be derived by applying 
Eq. \ref{inttra} to the $x$--derivative of the Euler equation (Malkov 1997; Malkov et 
al. 2000):
$$
\hat{U}(p) = \frac{R_{sub}-1}{R_{tot}} +
$$
$$
\frac{\nu}{p_{inj}}
\int_{p_{inj}}^{p_{max}} d\hat{p} \frac{\hat{p}}{\sqrt{\hat{p}^2+(mc)^2}}
\left[1+\frac{q(p)D(\hat{p})}{q(\hat{p})D(p)}\right]^{-1} \times
$$
\begin{equation}
\label{mal1}
\times \frac{\hat{U}(p_{inj})}{\hat{U}(\hat{p})}
\exp \left[ -\frac{3}{RR_{sub}}
\int_{p_{inj}}^{\hat{p}} \frac{d \ln p^\prime}{\hat{U}(p^\prime)}
\right],
\end{equation}
where $\nu$ is an injection parameter defined as
\begin{equation}
\nu = \frac{4\pi}{3} \frac{c}{\rho_0u_0^2} p_{inj}^4 f_0(p_{inj}),
\label{efficiency}
\end{equation}
and related to the compression factor by the following equation (Malkov 1997):
$$
\nu = p_{inj}(1-\frac{1}{R})
\Bigg\{
\int_{p_{inj}}^{p_{max}} dp \frac{p}{\sqrt{p^2+(mc)^2}}
\frac{\hat{U}(p_{inj})}{\hat{U}(p)}\times
$$
\begin{equation}
~~~~~\exp \left[ -\frac{3}{R_{sub}R} \int_{p_{inj}}^p 
\frac{d \ln p^\prime}{\hat{U}(p^\prime)}\right]\Bigg\}^{-1}.
\label{mal2}
\end{equation}
Here $R=R_{tot}/R_{sub}$ is the compression factor in the shock precursor. 
Eqs. \ref{eq:Rsub_Rtot}, \ref{mal1} and \ref{mal2} form a closed system that 
can be solved numerically.

Before showing the results it is worth noticing that the injection parameters 
$\eta$ and $\nu$ adopted in the two approaches are defined in two non equivalent 
ways. However, the relation between $\eta$ and $\nu$ can easily be found by using 
Eqs. \ref{eq:inje1} and \ref{efficiency} and can be written as:
\begin{equation}
\nu = \eta \left(\frac{p_{inj}c}{mu_0^2}\right) \frac{R_{tot}}{R_{sub}-1}.
\end{equation}
One may notice that this relation contains the compression factors $R_{tot}$ and 
$R_{sub}$, which are what we are searching for. This fact implies that three 
solutions characterized by the same value of $\eta$ may correspond to three 
distinct values of $\nu$. 

In order to compare the results of the two different approaches we consider a 
shock having Mach number $M_0 = 150$ and temperature at upstream infinity 
$T_0 = 10^4 K$.
We set the value of the injection and maximum momenta equal to $10^{-3} mc$ 
and $10^5  mc$ respectively and we adopt an efficiency $\eta = 10^{-4}$.
Using the approach proposed by Blasi (2002), we find three solutions, characterized
by the values of the compression factor $R \sim 15.3, 3.94, 1.05$. The last solution 
is the quasi-linear one, in which the precursor is very weak. We adopt now 
these three values for the precursor compression factor to solve the system of 
equations given by Eq. \ref{eq:Rsub_Rtot}, \ref{mal1}, and \ref{mal2}) for different 
choices of the diffusion coefficient. 

\begin{figure}
\resizebox{\hsize}{!}{\includegraphics{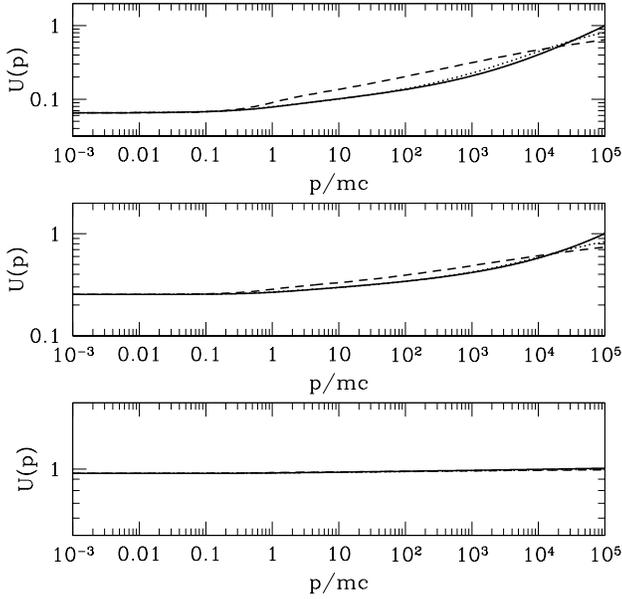}}
\caption[]{Velocity profile upstream of the shock as derived in this 
paper (solid line) and with the approach of Malkov (1997) with a 
Bohm diffusion (dotted line) and for a Kolmogorov diffusion (dashed
line).} 
\label{fig:malk_vel}
\end{figure}

In Fig. \ref{fig:malk_vel} we plot the velocity profiles for the three 
solutions, as derived with the method of Blasi (2002, 2004) and 
detailed above (solid line). In the figure $U(p)=u_p/u_0$ with $u_p$ 
defined through Eq. \ref{eq:up} for the method of Blasi (2002, 2004). 
It is easy to show that $U(p)$ is related to the $\hat{U}(p)$ through the 
relation $\hat U(p)+1/R_tot = U(p)$. The dotted and dashed lines are the 
results obtained with the calculation of Malkov (1997) with a Bohm and 
Kolmogorov diffusion respectively. For Bohm diffusion
the two approaches give very similar results. For Kolmogorov diffusion
the differencies are larger, as expected.

\begin{figure}
\resizebox{\hsize}{!}{\includegraphics{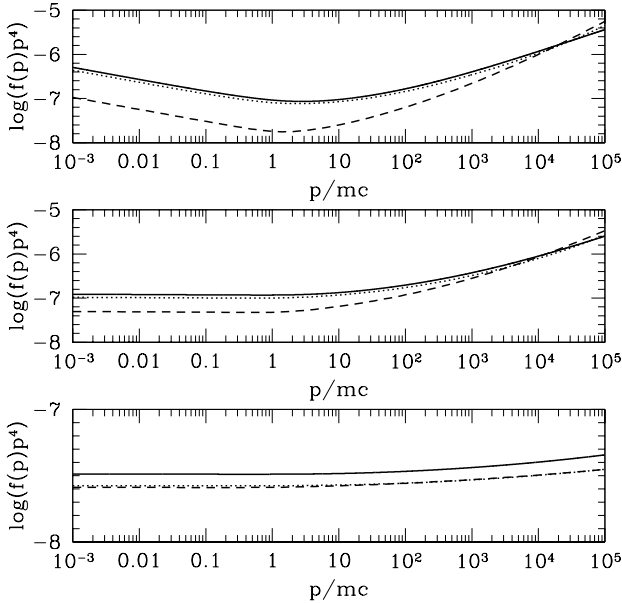}}
\caption[]{Spectrum of the accelerated particles as derived in this 
paper (solid line) and with the approach of Malkov (1997) with a 
Bohm diffusion (dotted line) and for a Kolmogorov diffusion (dashed
line).} 
\label{fig:malk_spec}
\end{figure}

In Fig. \ref{fig:malk_spec} we plot the spectra of the accelerated 
particles, as obtained in this paper (solid line) and for a Bohm (dotted 
line) and Kolmogorov (dashed line) diffusion coefficient, as derived by 
carrying out the calculation of Malkov (1997).

We recall that from the theoretical point of view the Bohm diffusion
is in fact what should be expected in the proximity of a shock if
the turbulence necessary for the acceleration is strong and generated
by the same cosmic rays that are being accelerated (Lagage \& Cesarsky 1983). 
In this perspective, we look at the results illustrated in this section 
as very encouraging in using the approach presented in Blasi (2002, 2004), 
since it is simple and at the same time accurate in reproducing the major 
physical aspects of particle acceleration at cosmic ray modified shocks. 

\section{A recipe for injection from the thermal pool}
\label{sec:injectionrecipe}

The presence of multiple solutions is typical of many non-linear
problems and should not be surprising from the mathematical point
of view. In terms of physical understanding however, multiple 
solutions may be disturbing. The typical situation that takes
place in nature when multiple solutions appear in the description 
of other non linear systems is that (at least) one of the solutions 
is unstable and the system {\it falls} in a stable solution when perturbed. 
The stable solutions are the only ones that are physically meaningful.
Some attempts to investigate the stability of cosmic ray modified
shock waves have been made by Mond \& Drury (1998) and Toptygin (1999), 
but all of them refer to the two-fluid models. A step forward is being 
carried out by Blasi \& Vietri (in preparation) in the context of kinetic 
models.

In addition to the stability, another issue that enters the physical 
description of our problem is the identification of possible processes
that determine some type of backreaction on the system. It may be 
expected that when some types of processes of self-regulation are included, 
the phenomenon of multiple solutions is reduced. In this section we investigate the 
type of reaction that takes place when a self-consistent, though 
simple, recipe for the injection of particles from the thermal pool is adopted. 
This recipe is similar to that proposed by Kang, Jones \& Gieseler
(2002) in terms of the underlying physical interpretation of the 
injection, but probably simpler in its implementation. 

For non-relativistic shocks, the distribution of particles downstream is 
quasi-isotropic, so that the flux of particles crossing the shock surface
from downstream to upstream can be written as
\begin{equation}
\Phi = - 2\pi ~ \int_{p_{min}}^\infty dp \int_{-1}^{-u_d/v(p)} 
d\mu\frac{f_{th}(p)}{4\pi}4\pi p^2 \left[u_d+v(p)\mu\right],
\label{eq:flux1}
\end{equation}
where $v(p)$ is the velocity of particles with momentum $p$ and $u_d$ is
the shock speed in the frame comoving with the downstream fluid. The 
term $u_d+v(p)\mu$ is the component along the direction perpendicular 
to the shock surface of the velocity of particles with momentum $p$ moving 
in the direction $\mu$. It follows that the flux of particles moving tangent to the shock
surface (namely with $\mu=-u_d/v(p)$) is zero. We recall that,
having in mind collisionless shocks, the typical thickness of the shock, 
$\lambda$, is the collision length associated with the magnetic interactions
that give rise to the formation of the discontinuity. Useless to say that
these interactions are all but well known, and at present the best we can do
is to attempt a phenomenological approach to take them into account, without
having to deal with their detailed physical understanding. It is however
worth recalling that many attempts have been made to tackle the problem
of injection at a more fundamental level (Malkov 1998; Malkov \& V\"{o}lk 1995; 
Malkov \& V\"{o}lk 1998). Here, we consider
the reasonable situation in which $\lambda\propto r_L^{th}$, where 
$r_L^{th}\propto p_{th}$ is the Larmor radius of the particles in the 
downstream fluid that carry most of the thermal energy, namely those
with momentum $1.5~p_{th}$ ($p_{th}=(2 m k_BT_2)^{1/2}$ here is the momentum 
of the particles in the thermal peak of the maxwellian distribution in the 
downstream plasma, having temperature $T_2$). 
We stress here the important point that the 
temperature of the downstream gas (and therefore $p_{th}$) is determined 
by the shock strength, which in the presence of accelerated particles, is 
affected by the pressure of the non-thermal component. In particular, 
the higher the efficiency of the shock as a particle accelerator, the 
weaker its efficiency in terms of heating of the background plasma
(see section \ref{sec:heating}).
 
For collisionless shocks, it is not clear whether the downstream plasma 
can actually be thermalized and the distribution function be a maxwellian. 
On the other hand, it is generally assumed that this is the case, so that 
in the following we consider the case in which the bulk of the background 
plasma is thermal and has a maxwellian spectrum at temperature $T$ given 
by the generalized Rankine-Hugoniot relations in the presence of accelerated 
particles (see section \ref{sec:nonlin}). 
For modified shocks, the points discussed above apply to the 
so-called subshock, where the injection of particles from the thermal pool 
is expected to take place. We recall that for strongly modified  
shocks the subshock is weak, and rather inefficient in the heating 
of the background plasma. 

\begin{figure*}
 \begin{center}
  \mbox{\epsfig{file=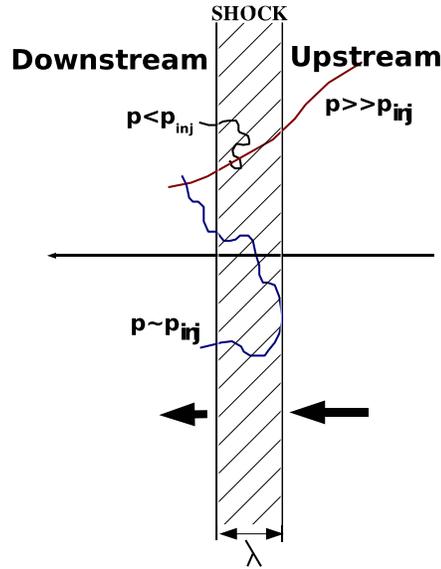,width=14.cm}}
  \caption{Graphic illustration of the structure of a collisionless shock and
of the basic idea underlying our injection recipe. }
 \end{center}
\label{fig:structure}
\end{figure*}

From Eq. \ref{eq:flux1} we get:
\begin{equation}
\Phi = \frac{1}{4} \int_{p_{min}}^\infty dp 4\pi p^2 f_{th}(p) 
\frac{(v(p)-u_d)^2}{v(p)},
\label{eq:flux2}
\end{equation}
where we assumed that the temperature downstream implies non-relativistic 
motion of the quasi-thermal particles ($p\approx m v(p)$). In Eq. \ref{eq:flux2} 
we write the minimum momentum in terms of a parameter $\alpha$, such that $\lambda 
= \alpha r_L^{th}$. With this formalism, the particles that can cross the shock 
surface are those that satisfy the condition:
\begin{equation}
p > p_{min} = 1.5~\alpha~p_{th}.
\end{equation}
The parameter $\alpha$ defines the thickness of the shock in units of the 
gyration radius of the bulk of the thermal particles.
Our recipe for injection is pictorially illustrated in Fig. 6:
thermal particles have a pathlength smaller than the shock thickness and cannot 
cross the shock surface, being advected downstream before the crossing occurs. 
Only particles with momentum sufficiently larger than the thermal momentum of the 
downstream particles can actually return upstream and be accelerated. 

In the following we will neglect the fluid speed $u_d$ compared with $v(p)$,
which is a good approximation if the injected particles are sufficiently more 
energetic than the thermal particles. This is done only to make the 
interpretation of the result simpler, but there is no technical difficulty 
in keeping the dependence of the results on $u_d$.

We introduce an effective injection momentum $p_{inj}=\xi p_{th}$ defined 
by the equation:
\begin{equation}
\Phi = \int_{\xi p_{th}}^\infty dp 4 \pi p^2 f_{th}(p) v(p),
\end{equation}
which in terms of dimensionless quantities, with $f_{th}(p)=
e^{-(p/p_{th})^2}$ reads:
\begin{equation}
\int_{1.5\alpha}^{\infty} dx x^3 e^{-x^2} =
4 \int_{\xi}^\infty dx x^3 e^{-x^2}.
\end{equation}
It is easy to show that $\xi\approx 2$ for $\alpha=1$ (half a Larmor 
rotation of the particles with momentum $1.5 p_{th}$ inside the thickness 
of the shock) and $\xi\approx 3.25$ for $\alpha=2$ (one full Larmor rotation 
of the particles with momentum $1.5 p_{th}$ inside the {\it thickness} of 
the shock. The fraction of particles at momentum $\xi$ times larger than 
the thermal one is $\sim 5\%$ for $\xi=2$ and $\sim 10^{-4}$ for 
$\xi=3.25$. The actual values of $\xi$ are expected to be somewhat higher
if the effect of advection with the downstream fluid is not neglected.
The sharp decrease in the fraction of {\it leaking} particles that may take
part in the acceleration process is due to the exponential behaviour of
the maxwellian at large momenta. Although the fraction of particles in the
maxwellian that get accelerated only depends on the parameter $\xi$ which
in turn is expected to keep the information about the microscopic structure
of the shock, the absolute number of and energy carried by these particles 
depend on the temperature of the downstream gas, which is an output of our 
calculation. This simple argument serves as an explanation of the physical 
reason why there is a nonlinear reaction on the system due to injection. 
If the parameter $\xi$ is assumed to be determined by the microphysics of 
the shock, and if we adopt our simple recipe to describe such microphysics, 
then the shock thickness is easily estimated once the temperature of the downstream 
gas is known, and the latter can be calculated from the modified Rankine-Hugoniot
relations. The parameter $\eta$ in Eq. \ref{eq:inje1} is no longer a free 
parameter, being related in a unique way to the parameter $\xi$ and to the
physical conditions at the subshock. The condition that fixes $\eta$ is 
that the total number of particles in the non-thermal spectrum equals the
number of particles in the maxwellian at momenta larger than $p_{inj}$. 
Due to the very strong dependence of the spectrum on the momentum for both 
the maxwellian and the power law at low momenta, the condition described above 
is very close to require the continuity of the distribution function, namely 
that $f_{th}(p_{inj}) = f_0(p_{inj})$. In the following we adopt this 
condition for the calculations.
This can be shown to imply the following expression for $\eta$:
\begin{equation}
\eta = \frac{4}{3\pi^{1/2}} (R_{sub}-1) \xi^3 e^{-\xi^2}.
\end{equation}
We recall that the compression factor at the subshock, $R_{sub}$, approaches
unity when the shock becomes cosmic ray dominated. This makes evident how 
the backreaction discussed above works: when the shock becomes increasingly more 
modified, the efficiency $\eta$ tends to decrease, limiting the amount of
energy that can be channelled in the non-thermal component. Although the 
recipe provided here is certainly far from representing the complexity of
the reality of injection of particles from the thermal pool, it may be
considered as a useful attempt to include the main physical aspects of
this phenomenon. 

\section{Self-consistent injection and multiple solutions}
\label{sec:singlesolutions}

\begin{figure}
\resizebox{\hsize}{!}{\includegraphics{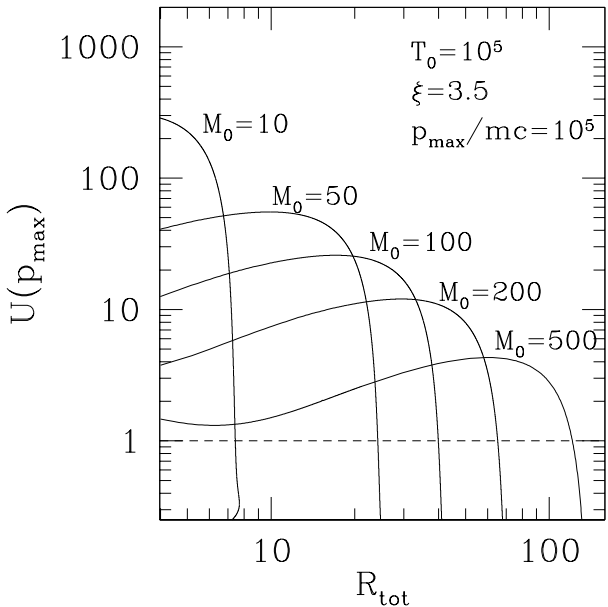}}
\caption{$U(p_{max}$ as a function of the total compression 
factor for $T_0=10^5$ K, $p_{max}=10^5~mc$ and $\xi=3$ for the 
Mach numbers indicated.} 
\label{fig:single}
\end{figure}

In this section we describe the role played by the injection recipe
discussed above for the appearance of multiple solutions. It can be 
expected that the phenomenon is somewhat reduced because, as discussed
in the previous section, the injection provides an efficient backreaction
mechanism on the shock as a particle accelerator. Indeed we find that 
the appearance of multiple solutions is drastically reduced, and that 
the phenomenon still exists only in regions of the parameter space 
which are very narrow and of limited physical interest. In the quantitative
calculations we use the value $\xi=3.5$ for the injection parameter, as
suggested by the simple estimate in Section \ref{sec:injectionrecipe} and
as suggested also in the numerical work of Kang \& Jones (1995). The 
dependence of the effect on the value of $\xi$ is discussed below. In 
Fig. \ref{fig:single} 
we illustrate the dramatic change in the physical picture by plotting 
$U(p_{max})$ as a function of $R_{tot}$ for $\xi=3.5$ and adopting the same 
values for the parameters as those used in obtaining Fig. 2. 
The efficiency $\eta$ is now calculated according with the 
recipe described in the previous section. It can be seen very clearly that 
when the Mach number of the shock is changed, there is a single solution 
(compare with Fig. 2 where multiple solutions where found for the same 
values of the parameters, but without thermal leakage).

\begin{figure*}
 \begin{center}
  \mbox{\epsfig{file=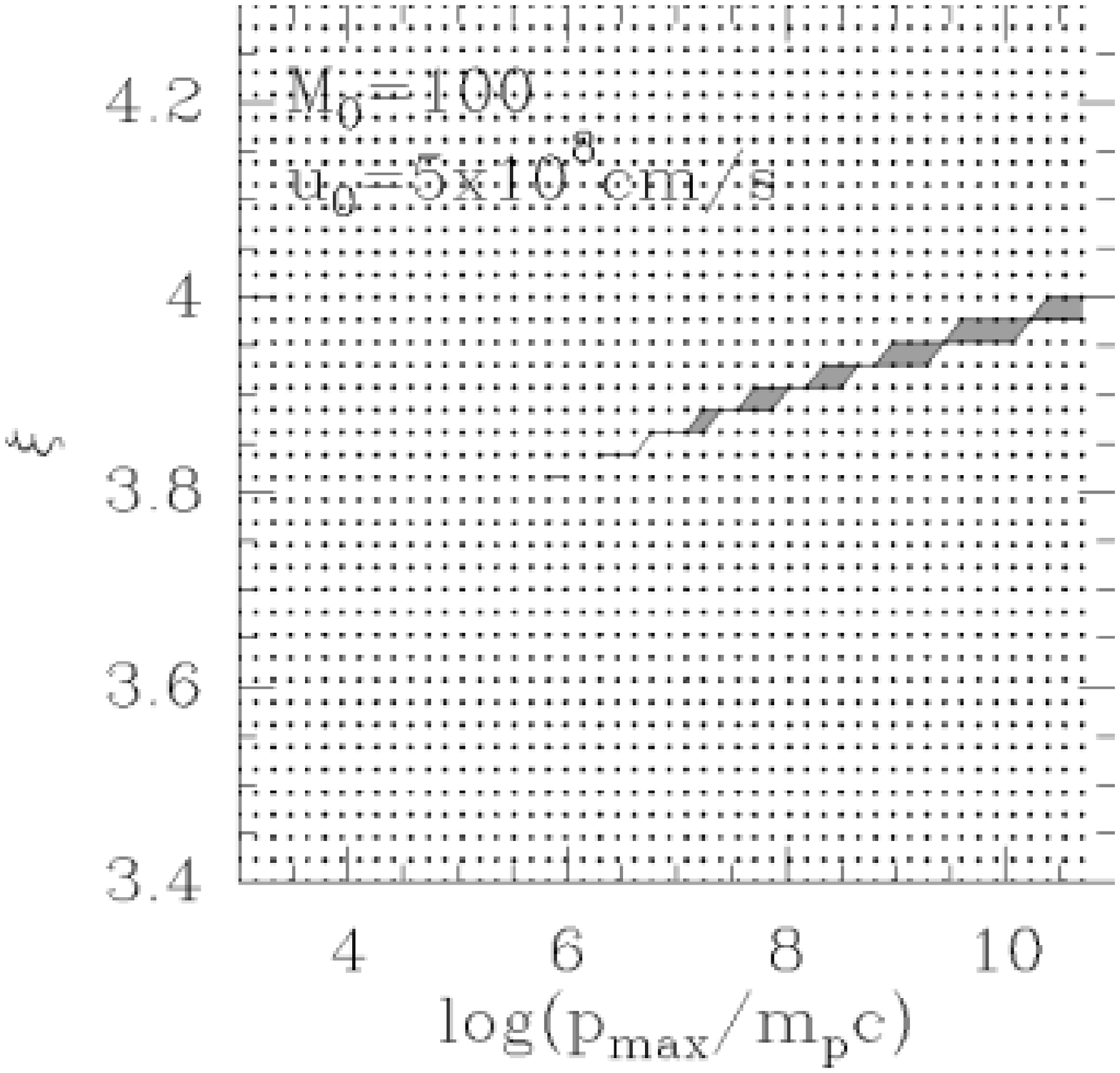,width=5.cm}}
  \mbox{\epsfig{file=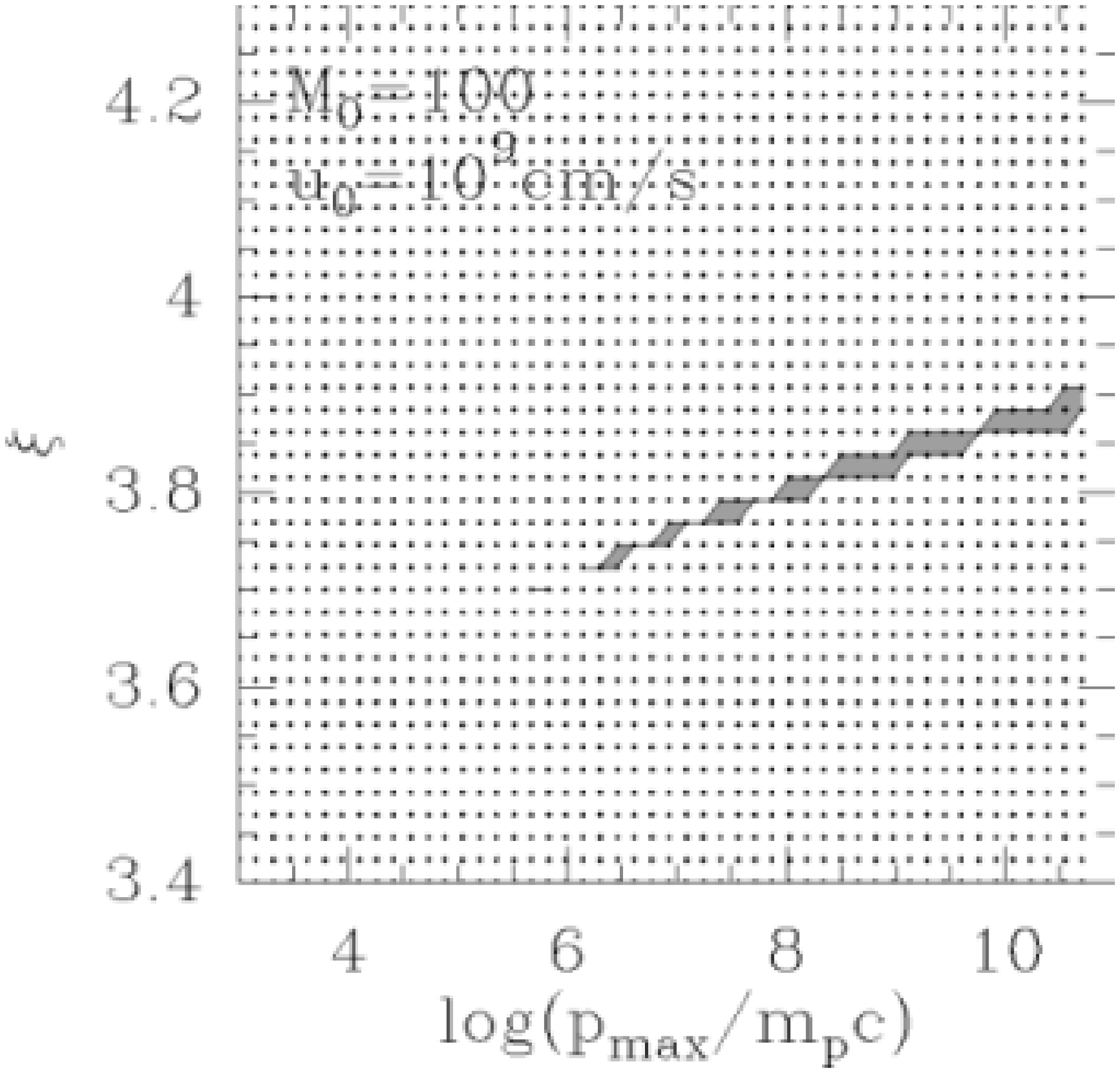,width=5.cm}}
  \mbox{\epsfig{file=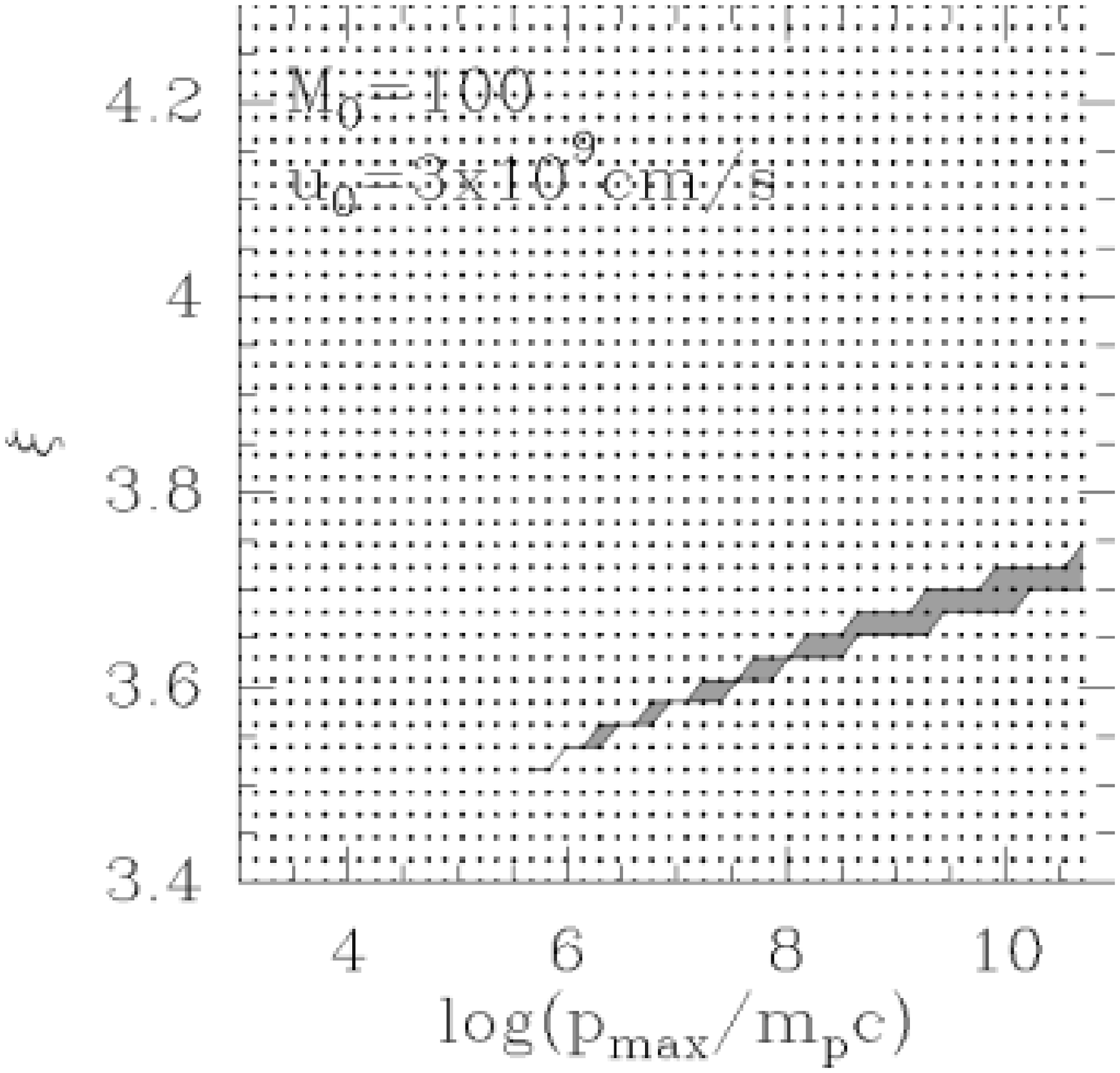,width=5.cm}}
  \mbox{\epsfig{file=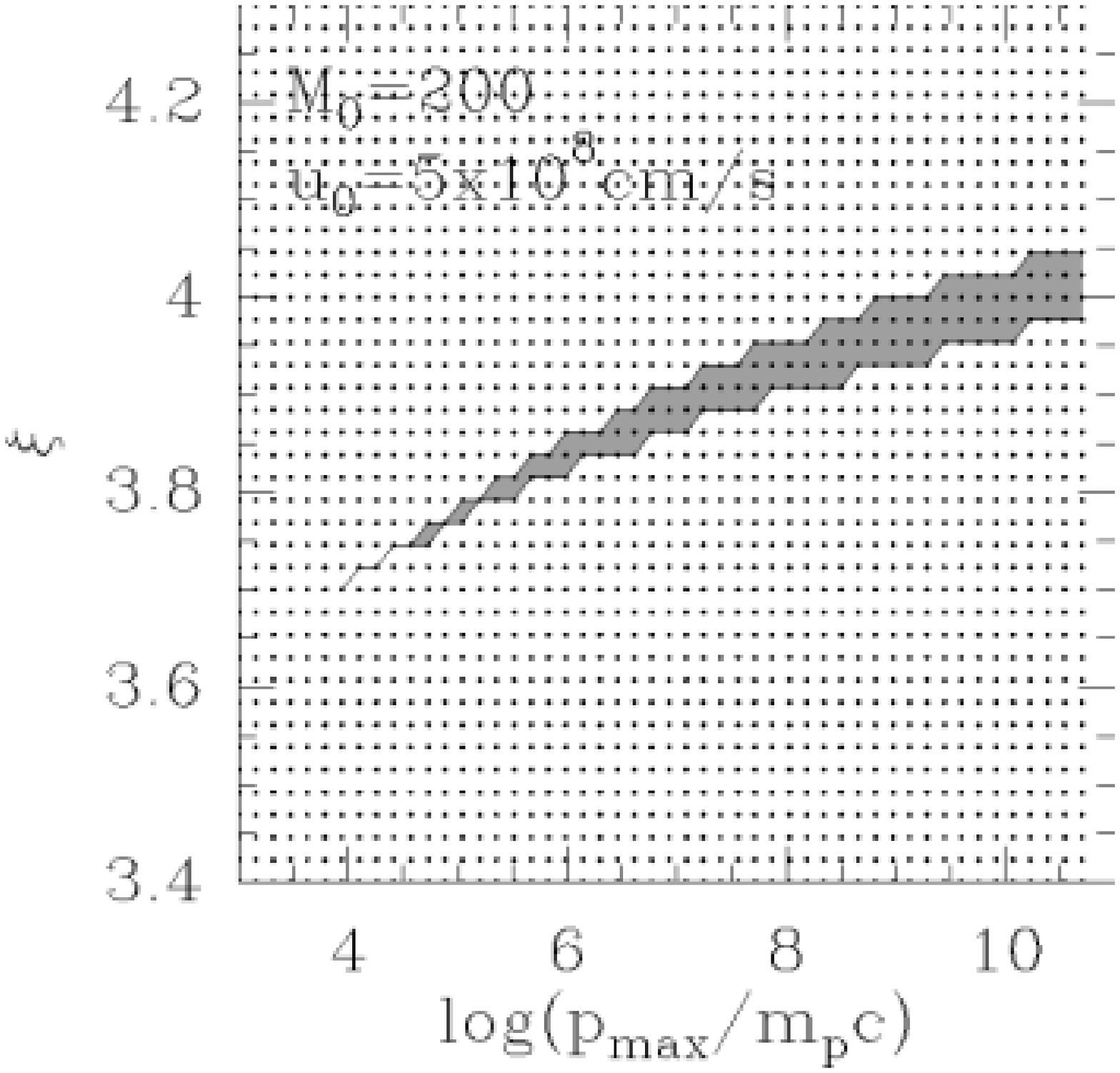,width=5.cm}}
  \mbox{\epsfig{file=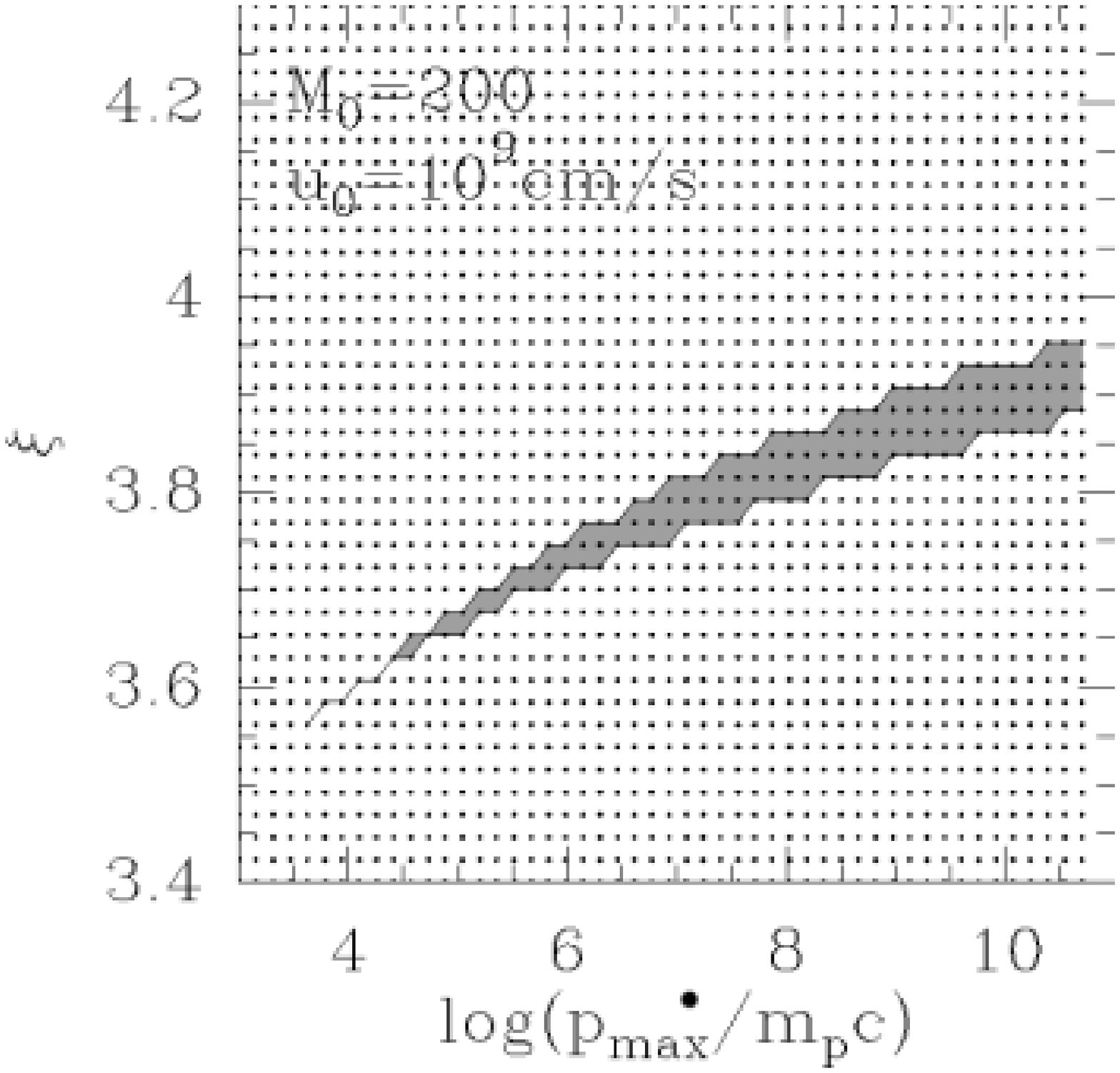,width=5.cm}}
  \mbox{\epsfig{file=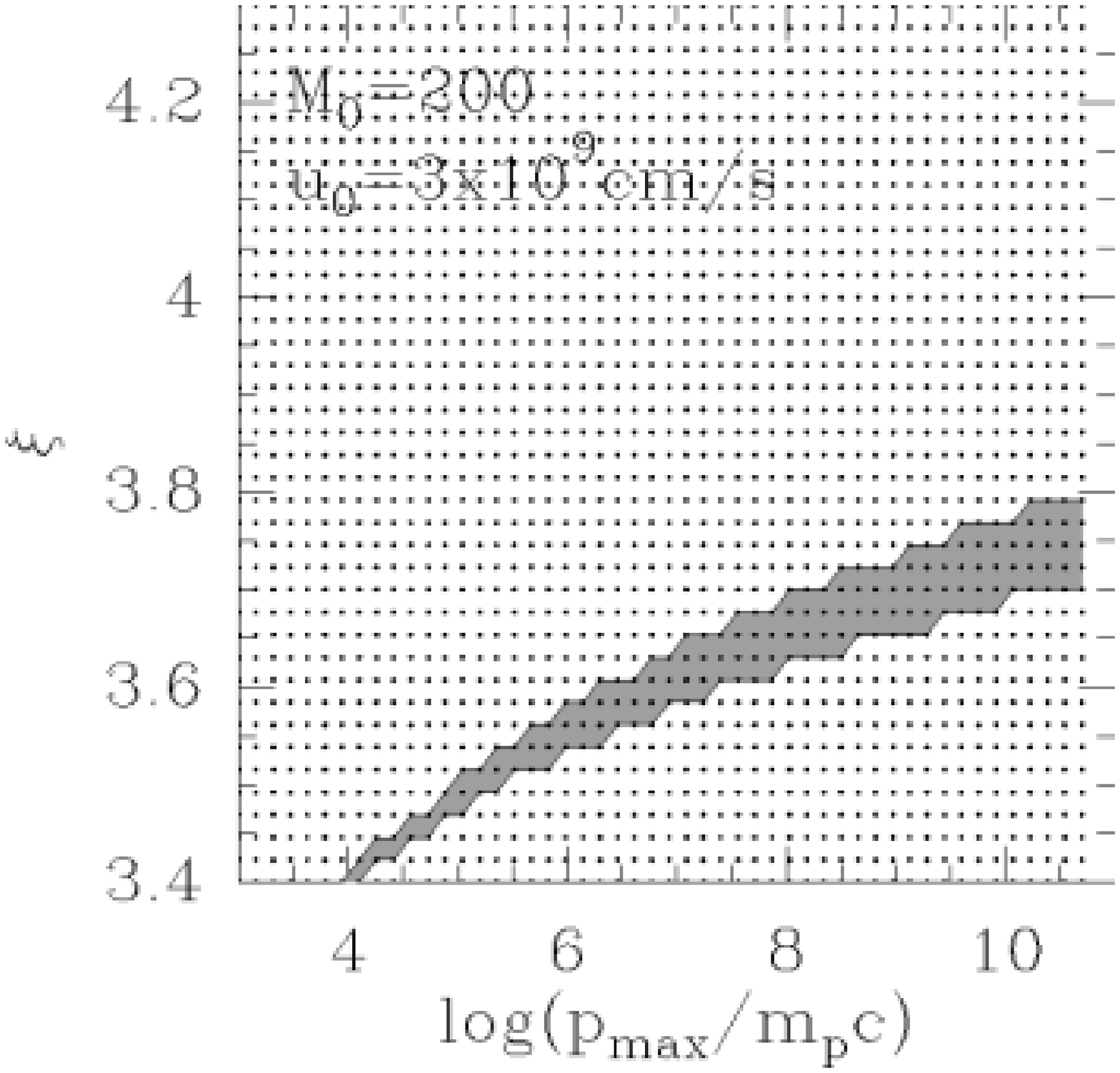,width=5.cm}}
  \mbox{\epsfig{file=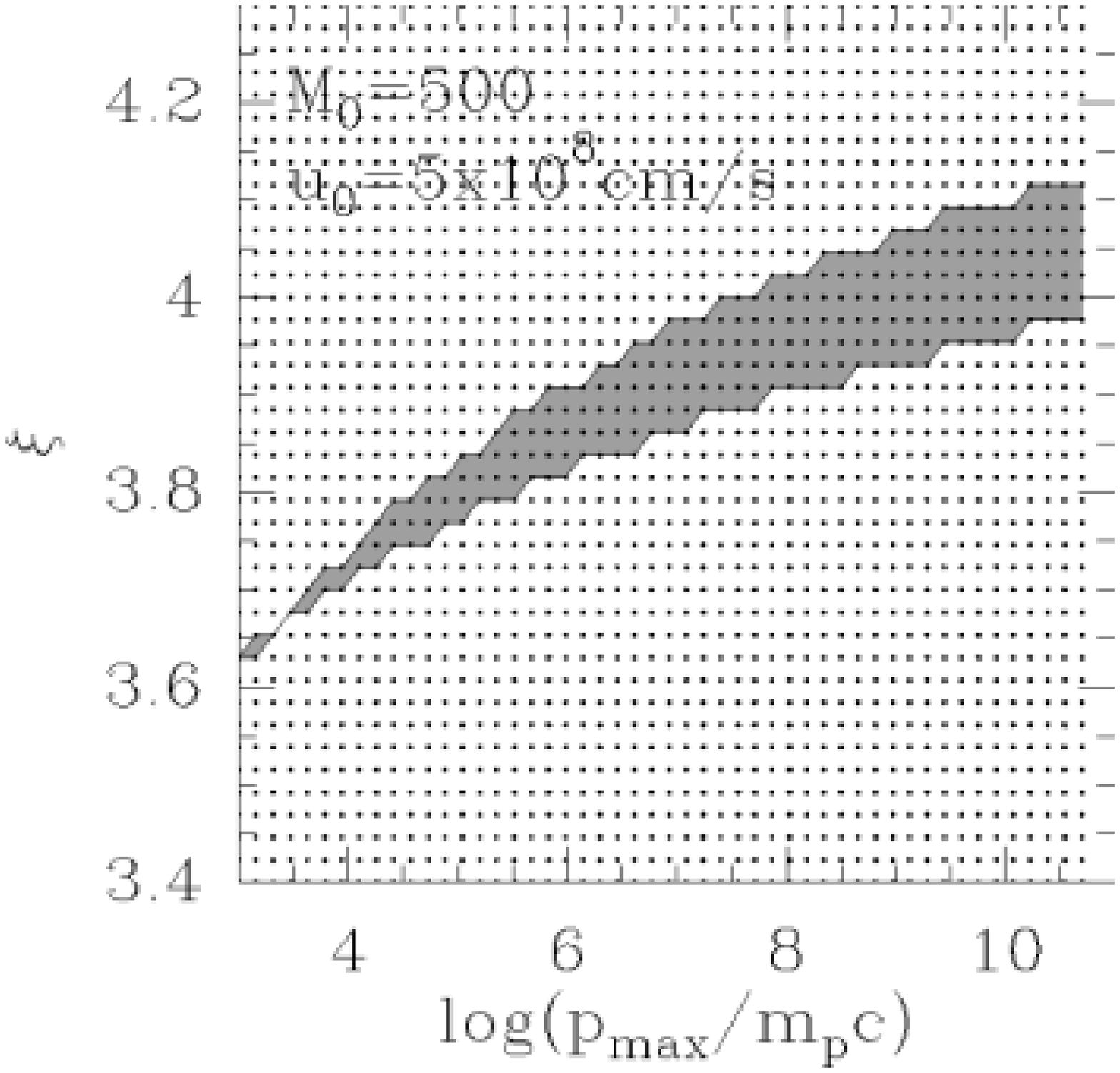,width=5.cm}}
  \mbox{\epsfig{file=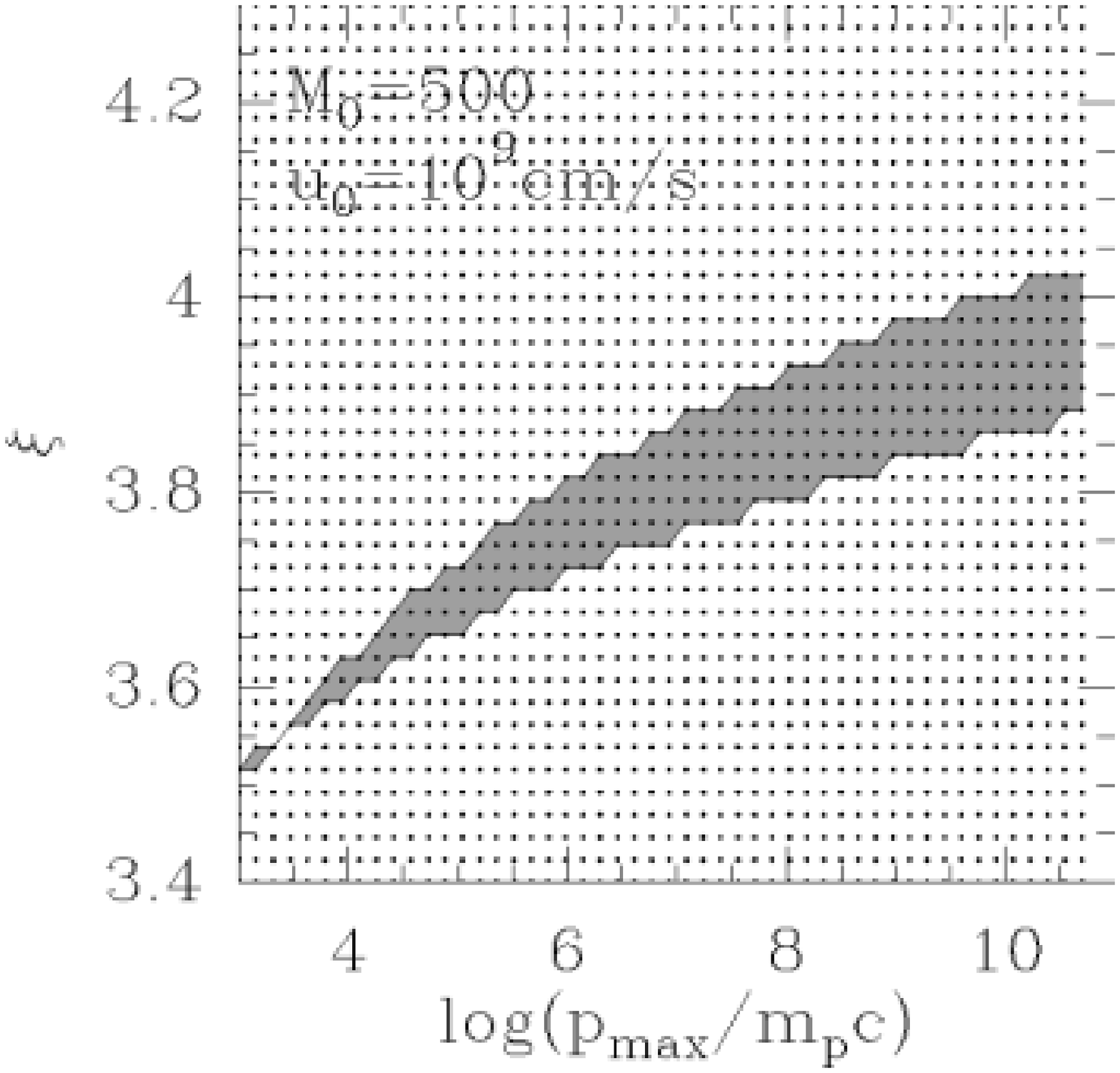,width=5.cm}}
  \mbox{\epsfig{file=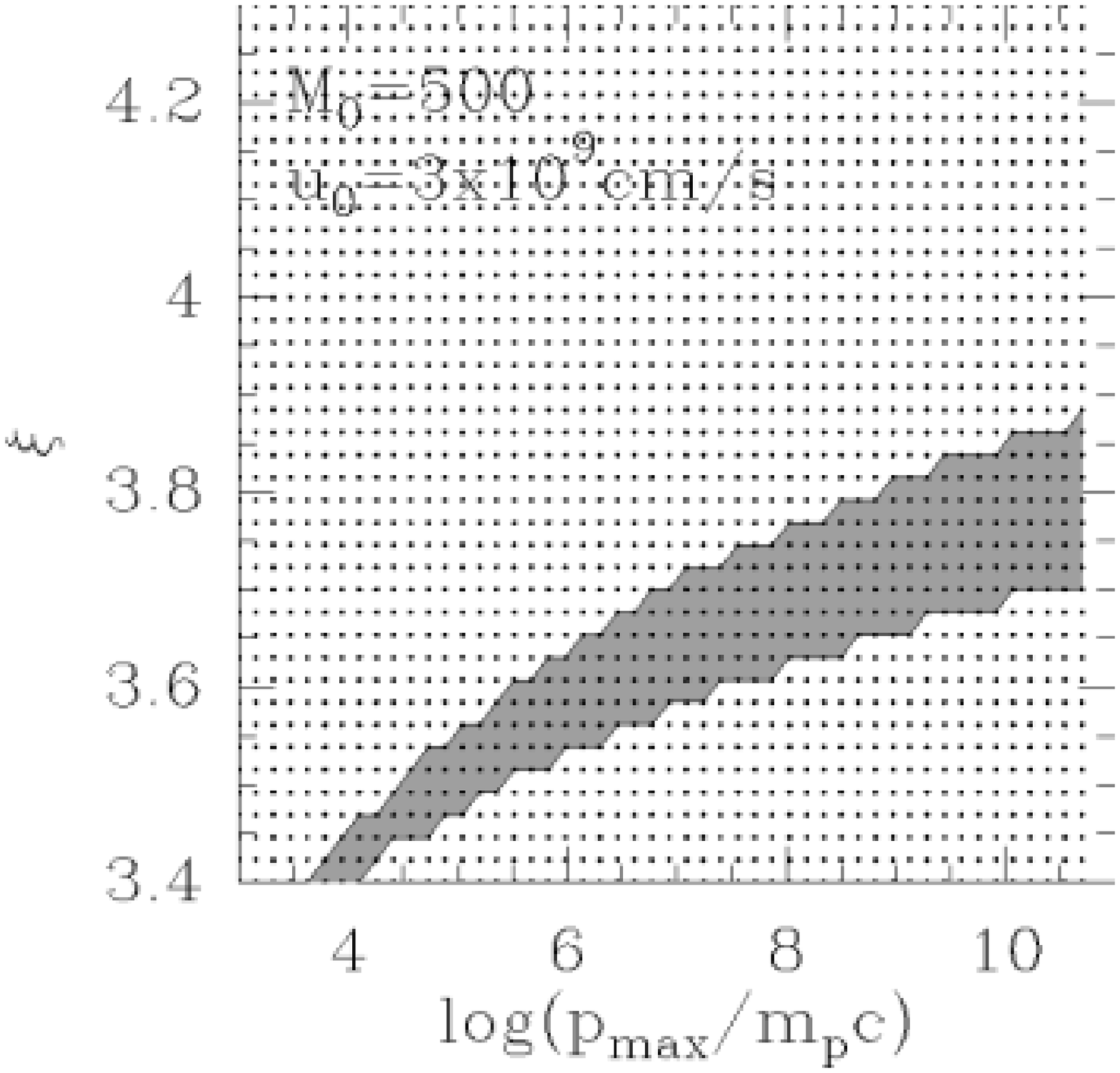,width=5.cm}}
  \caption{Parameter space for multiple solutions. The dark regions 
illustrate the regions of parameters for which multiple solutions are
still present.}
 \end{center}
\label{fig:parspace}
\end{figure*}

\begin{figure}
\resizebox{\hsize}{!}{\includegraphics{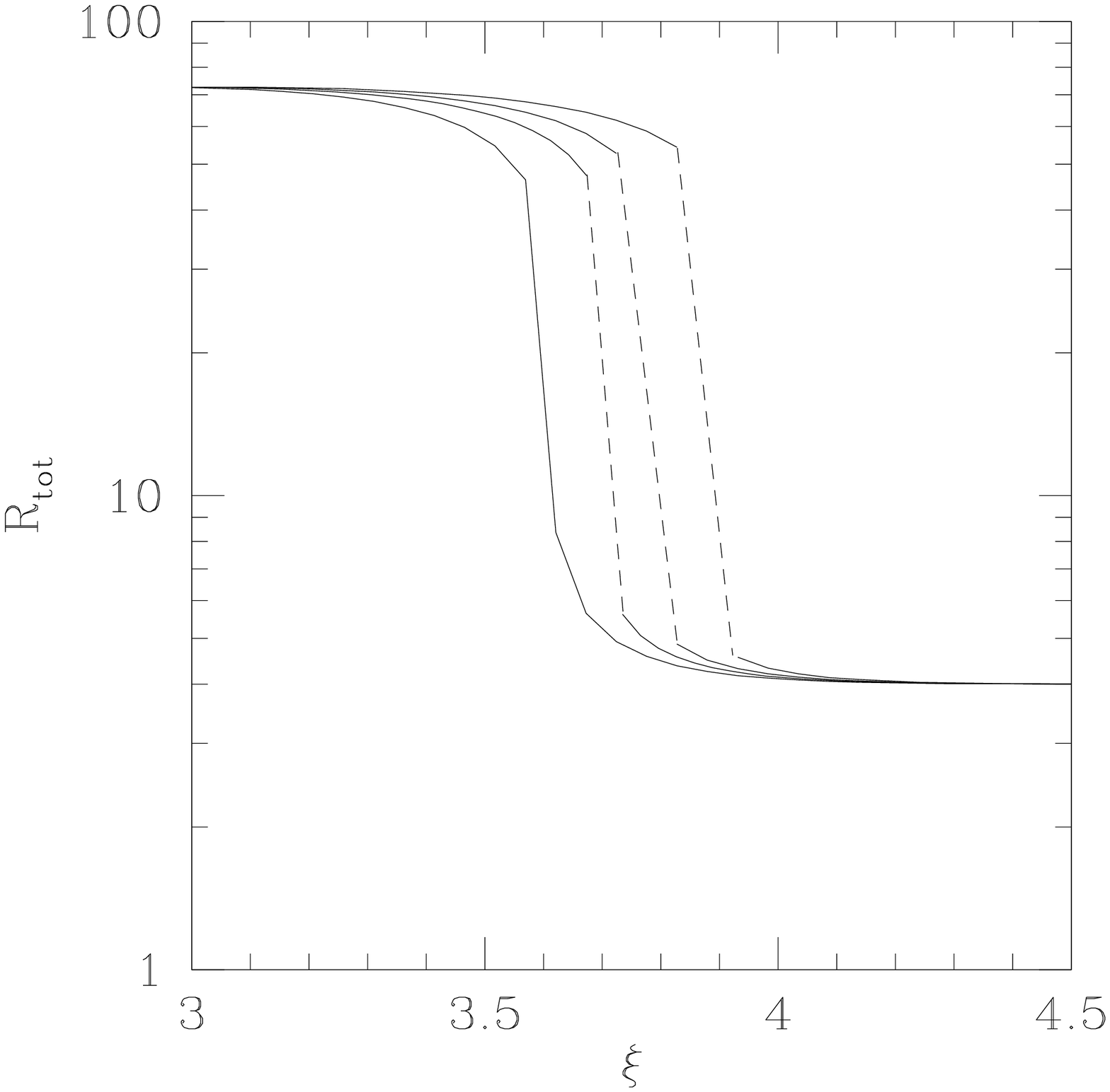}}
\caption[]{Dependence of $R_{tot}$ as a function of $\xi$ for 
$M_0=200$, $u_0=5\times 10^8\rm cm~s^{-1}$ and $p_{max}=
10^3,10^4,~10^5,~10^7~mc$ from left to right. The sharpness 
of the transition suggests that the small perturbations of the
parameters make the solution fall on one of the two sides.} 
\label{fig:transition}
\end{figure}

The appearance of multiple solutions can be investigated in the 
whole parameter space, in order to define the regions where the 
phenomenon appears, when it does. In Fig. 8
we highlight the regions where there are multiple solutions (dark regions)
in a plane $\xi-\log(p_{max})$, for different values of the Mach number 
of the shock. In most cases the dark regions are very narrow and cover
a region of values of $\xi$ which is rather high (small efficiency). 
In Fig. \ref{fig:transition} we plot the value of $R_{tot}$ as a function
of $\xi$ for $M_0=200$, $u_0=5\times 10^8 \rm cm~s^{-1}$ and $p_{max}=
10^3,10^4,~10^5,~10^7~mc$ from left to right. The line is continuous 
when there are no multiple solutions and dashed when multiple solutions 
appear. The dashed regions are, as stressed above, rather narrow.
For instance, for $p_{max}=10^4  mc$ there are multiple solutions
only for $3.67\leq\xi\leq3.7$. Any small perturbation of the system that
changes the values of $\xi$ at the percent level implies that the 
system shifts to one of the single solutions if it is sitting in 
the intermediate solution before the perturbation. 
The sharp transition between the strongly modified solution and the
quasi-linear solution when $\xi$ is increased suggests that the intermediate 
solution may be unstable, though a formal demonstration cannot be provided
here. In order to make sure that this is the case, a careful analysis of the 
stability is required, and will be presented in a forthcoming publication 
(Blasi \& Vietri, in preparation). 
On the other hand, a previous study, carried out in the context of the two-fluid 
models, showed that when multiple solutions are present, the solution with 
intermediate efficiency is in fact unstable to corrugations of the 
shock surface (Mond and Drury 1998).

\section{Escaping flux of accelerated particles}
\label{sec:escape}

It is rather remarkable that the kinetic model of Blasi (2002, 2004)
does not require explicitely the use of the equation for energy flux
conservation. However, once the solution of the kinetic problem has 
been found, the equation for conservation of the energy flux provides 
very useful information, as we show below. The equation can be written 
in the following form:
$$
\frac{1}{2} \rho_2 u_2^3 +\frac{\gamma_g}{\gamma_g-1} P_{g,2} u_2 +
\frac{\gamma_c}{\gamma_c-1} P_{c,2} u_2 =
$$
\begin{equation} 
~~~~~~~~~~
\frac{1}{2} \rho_0 u_0^3 +\frac{\gamma_g}{\gamma_g-1} P_{g,0} u_0 - F_E,
\label{eq:energy}
\end{equation}
where $F_E$ is the flux of particles escaping at the maximum momentum from
the upstream section of the fluid (Berezhko \& Ellison, 1999). Notice that 
this term is usually neglected 
in the linear approach to particle acceleration at shock waves because the
spectra are steep enough that, in most cases, we can neglect the
flux of particles leaving the system at the maximum momentum. The fact that
particles leave the system make the upstream fluid behave as a radiative 
fluid, and makes it more compressible. This is a crucial consequence
of particle acceleration at modified shocks, and is shown here to be a 
natural consequence of energy conservation. 

\begin{figure*}
 \begin{center}
  \mbox{\epsfig{file=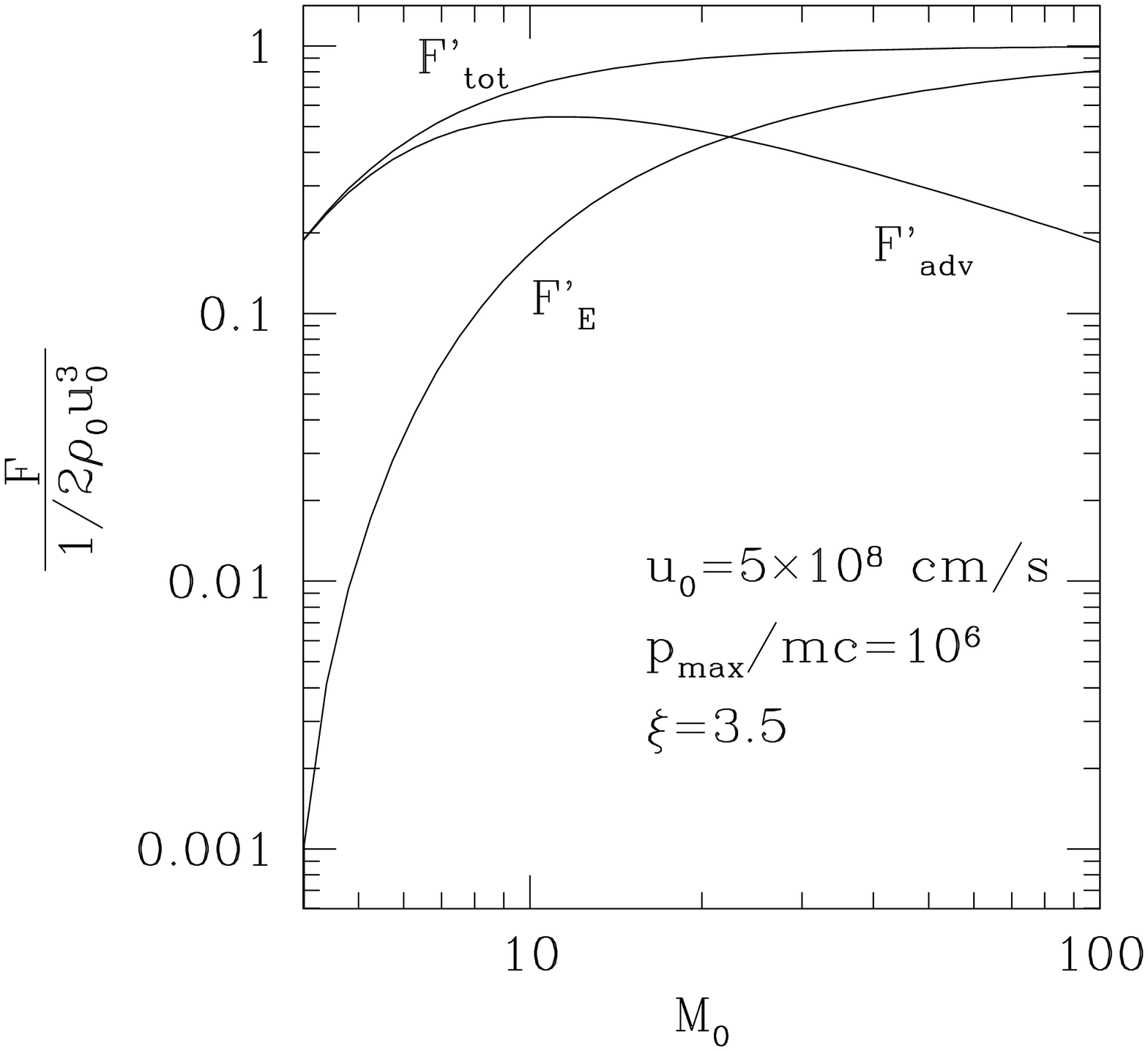 ,width=8.cm}}
  \mbox{\epsfig{file=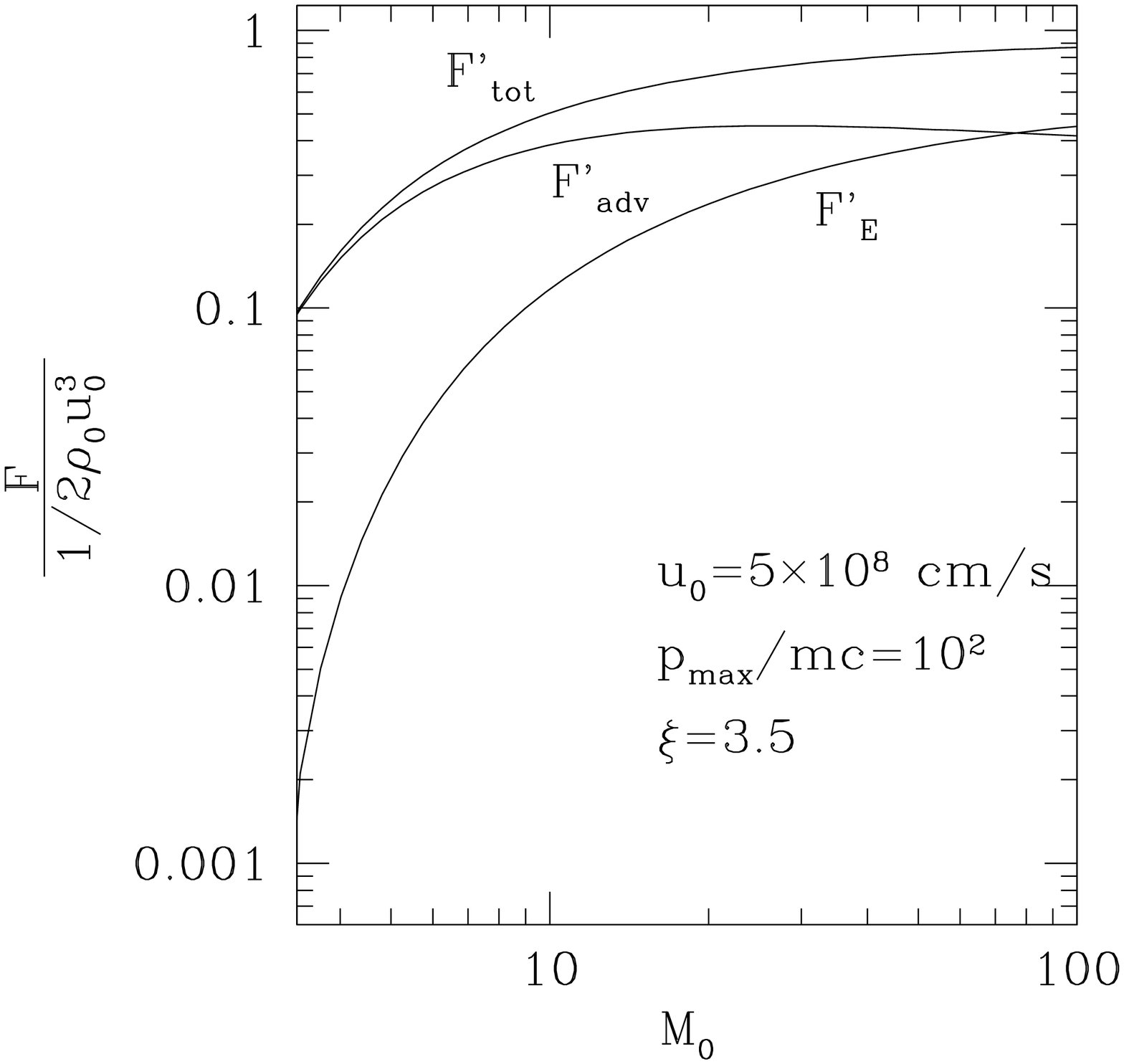 ,width=8.cm}}
\caption[]{Escaping flux ($F_E'$), advected flux ($F_{adv}'$) and the sum 
of the two ($F_{tot}'$) normalized to the incoming flux $(1/2)\rho_0 u_0^3$, 
as functions of the Mach number at upstream infinity $M_0$. Left panel:
$u_0=5\times 10^8~\rm cm~s^{-1}$, $p_{max}=10^6~ mc$ and $\xi=3.5$.
Right Panel: $u_0=5\times 10^8~\rm cm~s^{-1}$, $p_{max}=10^2~ mc$ 
and $\xi=3.5$.}
\end{center}
\label{fig:escape}
\end{figure*}

In Eq. \ref{eq:energy} we can divide all terms by $(1/2)\rho_0 u_0^3$ and
calculate the normalized escaping flux:
$$
F_E' = 1 - \frac{1}{R_{tot}^2} +\frac{2}{M_0^2 (\gamma_g-1)}
-\frac{2}{R_{tot}}\frac{\gamma_g}{\gamma_g-1}\frac{P_{g,2}}{\rho_0 u_0^2}
$$
\begin{equation}
~~~~~
-\frac{2}{R_{tot}}\frac{\gamma_c}{\gamma_c-1}\frac{P_{c,2}}{\rho_0 u_0^2}.
\label{eq:energy_norm}
\end{equation}
From momentum conservation at the subshock we also have:
\begin{equation}
\frac{P_{c,2}}{\rho_0 u_0^2} = \frac{R_{sub}}{R_{tot}} -
\frac{1}{R_{tot}} + \frac{1}{\gamma_g M_0^2} \left( 
\frac{R_{sub}}{R_{tot}}\right)^{-\gamma_g},
\end{equation}
so that the escaping flux only depends upon the {\it environment}
parameters (for instance the Mach number at upstream infinity) and
the compression parameter $R_{sub}$ which is part of the solution.
Note also that the adiabatic index for cosmic rays, $\gamma_c$, is
here calculated self-consistently as:
\begin{equation}
\gamma_c = 1 + \frac{P_c}{E_c} = 1 + 
\frac{\frac{1}{3}\int_{p_{inj}}^{p_{max}} d p 4 \pi p^3 v(p) f_0(p)}
{\int_{p_{inj}}^{p_{max}} d p 4 \pi p^2 f_0(p) \epsilon(p)},
\end{equation}
where $E_c$ is the energy density in the form of accelerated particles
and $\epsilon(p)$ is the kinetic energy of a particle with momentum $p$.
It can be easily seen that $\gamma_c\to 4/3$ when the energy budget 
is dominated by the particles with $p\sim p_{max}$ (namely for strongly
modified shocks) and $\gamma_c\to 5/3$ for weakly modified shocks.
In Eq. \ref{eq:energy_norm} the term $F_{adv}'=\frac{2}{R_{tot}}
\frac{\gamma_c}{\gamma_c-1}\frac{P_{c,2}}{\rho_0 u_0^2}$ is clearly the 
fraction of flux which is advected downstream with the fluid. 

In Fig. 10 we plot the escaping flux ($F_E'$), the 
advected flux ($F_{adv}'$) and the sum of the two ($F_{tot}'$) normalized
to the incoming flux $(1/2)\rho_0 u_0^3$, as functions of the Mach number
at upstream infinity $M_0$. Here we used $u_0=5\times 10^8~\rm cm~s^{-1}$,
and $\xi=3.5$, while the maximum momentum has been chosen as 
$p_{max}=10^6~ mc$ in the left panel and $p_{max}=10^2~ mc$ in the 
right panel. Several comments are in order: 
\begin{itemize}
\item[1)] At low Mach numbers the escaping flux is inessential, as one would 
expect for a weakly modified shock. We recall that the escaping flux is due to
the particles with momentum $p_{max}$ leaving the system from upstream
infinity. For a weakly modified shock at low Mach number the spectrum is 
steeper than $E^{-2}$, so that the energy carried by the highest energy 
particles is a small fraction of the total. 

\item[2)] At large Mach numbers the shock becomes increasingly more 
cosmic ray dominated, and for the cases at hand the total efficiency
gets very close to unity, meaning that the shock behaves as an extremely 
efficient accelerator. At Mach numbers around 4 on the other hand the 
total efficiency is around $20\%$ for $p_{max}=10^6~ mc$ and $\sim 10\%$
for $p_{max}=10^2~ mc$, dropping fast below Mach number 4.
Clearly the efficiency would be higher in this region for lower values
of the parameter $\xi$.

\item[3)] Despite the fact that the total efficiency of the shock as
a particle accelerator is close to unity at large Mach numbers, the 
fraction of the incoming energy which is actually advected toward 
downstream infinity is only $\sim 20\%$ at $M_0\approx 100$ for 
$p_{max}=10^6~ mc$. Most of the enegy flux in this case is in fact 
in the form of energy escaping from upstream infinity at the highest 
momentum $p_{max}$. For $p_{max}=10^2~ mc$ the normalized advected flux 
roughly saturates at $\sim 40\%$ and is comparable with the escape
flux at the same Mach number. For a distant observer these escaping 
particles would have a spectrum close to a delta function around $p_{max}$.
\end{itemize}

\section{Shock heating in the presence of efficient particle acceleration}
\label{sec:heating}

\begin{figure}
\resizebox{\hsize}{!}{\includegraphics{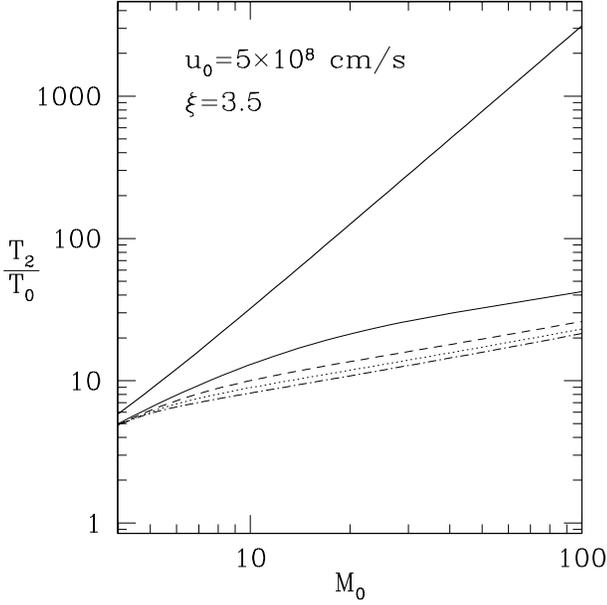}}
\caption[]{Temperature jump $T_2/T_0$ for $p_{max}/mc=10^3$ (thin 
solid line), $p_{max}/mc=10^5$ (dashed line), $p_{max}/mc=10^7$ 
(dotted line) and $p_{max}/mc=5\times 10^{10}$ (dash-dotted line). 
The thick solid line shows the jump for ordinary shocks.} 
\label{fig:jump}
\end{figure}

Energy conservation has the natural consequence that a smaller fraction
of the kinetic energy of the fluid is converted into thermal energy of the 
downstream plasma in cosmic ray modified shocks, compared with the case 
of ordinary shocks. The reduction of the heating at nonlinear shock
waves is fully confirmed by our calculation in the context of the 
injection recipe introduced in section \ref{sec:injectionrecipe}. In 
Fig. \ref{fig:jump} we plot the temperature jump between downstream 
infinity (at temperature $T_2$) and upstream infinity (at temperature 
$T_0$). The thick solid line is the jump as predicted by the standard
Rankine-Hugoniot relations without cosmic rays. The other lines 
represent the temperature jump at cosmic ray modified shocks with
$p_{max}/mc=10^3$ (thin solid line), $p_{max}/mc=10^5$ (dashed line), 
$p_{max}/mc=10^7$ (dotted line) and $p_{max}/mc=5\times 10^{10}$ 
(dash-dotted line).

Such a drastic reduction of the downstream temperature is expected
to reflect directly in the thermal emission of the downstream gas
in those environments in which collisions are relevant. Note that
for strongly modified shocks the compression factor between 
upstream infinity and downstream are much larger than for ordinary 
shocks, so that the downstream turns out to be denser but colder
than in the linear case. The missing energy ends up in the form
of accelerated particles. 

\begin{figure}
\resizebox{\hsize}{!}{\includegraphics{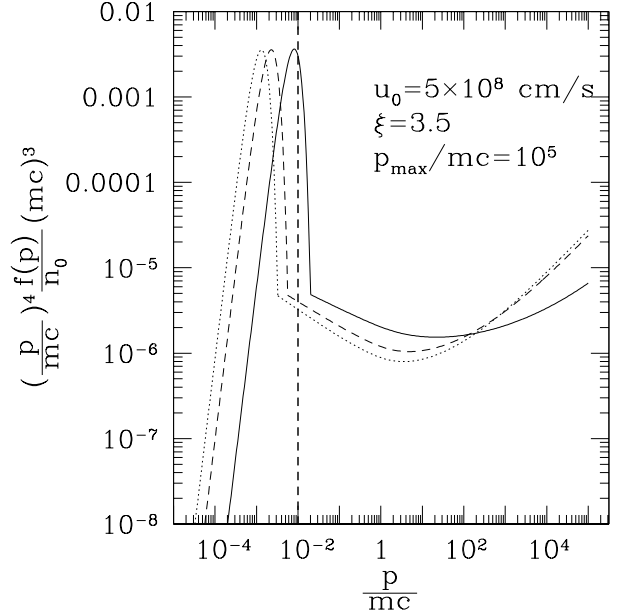}}
\caption[]{Particle Spectra (thermal plus non-thermal) for $M_0=10$
(solid line), $M_0=50$ (dashed line) and $M_0=100$ (dotted line). The 
vertical dashed line represents the position of the thermal peak for 
ordinary shocks (it is almost independent of the Mach number for large 
Mach numbers). }
\label{fig:spec}
\end{figure}

The effect of suppression of the heating in cosmic ray modified 
shocks also appears in the spectra of the particles (thermal plus 
non-thermal) in the shock vicinity. In Fig. \ref{fig:spec} we
show these spectra (including the maxwellian thermal bump) for 
$u_0=5\times 10^8~\rm cm~s^{-1}$, $\xi=3.5$ and $p_{max}/mc = 10^5$.
The vertical dashed line shows the position of the thermal peak as expected 
in the absence of accelerated particles. In fact this position should depend
on the Mach number, but the dependence is very weak for large Mach numbers.
The positions of the thermal peaks clearly show the effect of cooler 
downstream gases for modified shocks. At the same time, the effect is 
accompanied by increasingly more modified spectra of accelerated particles, 
with most of the energy pushed toward the highest momenta. 

\section{Conclusions}
\label{sec:conclusions}

The efficiency of the first order Fermi acceleration at shock fronts
depends in a crucial way upon details of the mechanism that determines
the injection of a small fraction of the particles from the thermal pool
to the {\it acceleration box}. In reality the processes of formation of
a collisionless shock wave, of plasma heating and particle acceleration 
are expected to be all parts of the same problem, though on different 
spatial scales.
We hide our lack of knowledge of the microphysics of the shock structure
in a simple recipe for injection, in which the particles that take part
in the acceleration process are those that have momentum larger by a factor
$\xi$ than the momentum of the thermal particles in the downstream
fluid. This is motivated by the fact that for collisionless shocks the 
thickness of the shock surface is determined by the gyromotion of the 
bulk of the thermal particles. We estimated that $\xi\sim 2-4$. This 
recipe implies that the fraction of particles that get accelerated is 
rather small ($0.01-10^{-5}$). 

We implemented this recipe in a calculation of the non-linear reaction
of cosmic rays on the shock structure proposed by Blasi (2002, 2004).
Similarly to other models, also this approach shows the appearance of 
multiple solutions, for a wide 
choice of the parameters of the problem. When the simple model of particle
injection is used, this phenomenon is drastically reduced: the multiple
solutions disappear for most of the parameter space, and when they appear 
they look as narrow strips in the parameter space, at the boundary between 
unmodified and modified shocks. We argued that this result suggests that 
the narrow regions may signal the transition between two stable 
solutions, although this needs further confirmation through detailed analyses
of the stability of the solutions. This interpretation seems to be 
supported in part by the calculations of Mond \& Drury (1998), that 
showed that when three solutions are present, the intermediate one is
indeed unstable for small corrugations of the shock front. This 
calculations was however performed in the context of a two-fluid model,
while an investigation of the stability for kinetic models is still
lacking. 

We find that the phenomenology of the particle acceleration at modified shocks 
is characterized by three main features:

\begin{itemize}
\item[1)] The modification of the shock increases with the Mach number
of the fluid. For low Mach numbers the quasi-linear solution is recovered,
but departures from it are evident already at relatively low Mach numbers. 
The modification of the spectra manifests itself with a hardening at high 
momenta and a softening at low momenta. The $p^4 f_0 (p)$ shows a characteristic
dip at intermediate momenta, typically around $p/mc \sim 1-100$ (for very
large values of $p_{max}$, the dip can be found at even larger momenta,
which is of interest for the acceleration of ultra-high energy cosmic rays).

\item[2)] The total efficiency for particle acceleration saturates at large
Mach numbers at a number of order unity. However, as shown in Fig. 10, the 
largest fraction of the energy is not advected downstream but rather escapes 
from upstream infinity at the maximum momentum. This effect was also 
discussed in the context of the simple model by Berezhko \& Ellison (1999).

\item[3)] The high efficiency for particle acceleration reflects in a 
reduced ability of cosmic ray modified shocks in the heating of the 
background plasma. This effect is at the very basis of the backreaction 
introduced by the injection recipe on the acceleration process, and
determines the saturation of the total efficiency for large Mach numbers. 
The heating suppression is shown in Fig. \ref{fig:jump} and in Fig. 
\ref{fig:spec}. 
\end{itemize}

\section*{Acknowledgements}
PB is grateful to E. Amato, L.O'C. Drury, D. Ellison and M. Vietri for 
many useful discussions. The research of PB was partially funded though 
COFIN-2004 at the Arcetri Astrophysical Observatory.
SG gratefully acknowledges support from the Alexander von Humboldt
Foundation. He also thanks Stephane Barland for useful discussions on
nonlinear systems.


\begin{thebibliography}{99}
 
\bibitem{ax_l_mk82}
Axford, W.I., Leer, E., and McKenzie, J.F., 1982, A\&A 111, 317

\bibitem{bell78}
Bell, A.R., 1978, MNRAS 182, 443

\bibitem{bell87}
Bell, A.R., 1987, MNRAS 225, 615

\bibitem{bell2004}
Bell, A.R., 2004, MNRAS, 353, 550

\bibitem{berezhko96}
Berezhko, E.G., 1996, Astropart. Phys. 5, 367

\bibitem{simple}
Berezhko, E.G., and Ellison, D.C., 1999, ApJ 526, 385

\bibitem{bk88}
Berezhko, E.G. and Krimsky, G.F., 1988, Soviet. Phys.-Uspekhi 12, 155

\bibitem{berezhko95}
Berezhko, E.G., Ksenofontov, L.T., and Yelshin, V.K., 1995, Nucl. Phys. B 
(Proc. Suppl.) 39A, 171

\bibitem{berezhko94}
Berezhko, E.G., Yelshin, V.K., and Ksenofontov, L.T., 1994, Astropart. Phys. 2, 215

\bibitem{blandford80}
Blandford, R.D., 1980, Astrophys. J. 238, 410

\bibitem{be87} 
Blandford, R.D. and Eichler, D., 1987, Phys. Rep. 154, 1

\bibitem{bo78}
Blandford, R.D. and Ostriker, J.P., 1978, ApJ Lett. 221, 29

\bibitem{blasi1}
Blasi, P., 2002, Astropart. Phys. 16, 429

\bibitem{blasi2}
Blasi, P., 2004, Astropart. Phys. 21, 45

\bibitem{drury83}
Drury, L .O'C., 1983, Rep. Prog. Phys. 46, 973

\bibitem{dr_ax_su82}
Drury, L.O'C, Axford, W.I. and Summers, D., 1982, MNRAS 198, 833

\bibitem{dr_v80}
Drury, L.O'C and V\"{o}lk, H.J., 1980, Proc. IAU Symp. 94, 363

\bibitem{dr_v81}
Drury, L.O'C and V\"{o}lk, H.J., 1981, ApJ 248, 344

\bibitem{ddv94}
Duffy, P., Drury, L.O'C. and V\"{o}lk, H.J., 1994, A\&A 291, 613

\bibitem{eich84a}
Eichler, D., 1984, ApJ 277, 429

\bibitem{eich85}
Eichler, D, 1985, ApJ 294, 40

\bibitem{ebj95}
Ellison, D.C., Baring, M.G., and Jones, F.C., 1995, ApJ 453, 873

\bibitem{ebj96}
Ellison, D.C., Baring, M.G., and Jones, F.C., 1996, ApJ 473, 1029

\bibitem{eich84b}
Ellison, D., and Eichler, D., 1984, ApJ 286, 691

\bibitem{elleich85}
Ellison, D.C., and Eichler, E., 1985, Phys. Rev. Lett. 55, 2735

\bibitem{elli90}
Ellison, D.C., M\"{o}bius, E., and Paschmann, G., 1990, ApJ 352, 376

\bibitem{gieseler}
Gieseler, U.D.J., Jones, T.W., and Kang, H., 2000, A\&A 364, 911

\bibitem{je91}
Jones, F.C. and Ellison, D.C., 1991, Space Sci. Rev. 58, 259

\bibitem{lagage}
Lagage, P. O., and Cesarsky, C. J., 1983, A\&A, 118, 223

\bibitem{kj95}
Kang, H. and Jones, T. W., 1995, ApJ 447, 944

\bibitem{kj97}
Kang, H., and Jones, T.W., 1997, ApJ 476, 875

\bibitem{kj2005}
Kang, H., and Jones, T.W., 2005, ApJ 620, 44

\bibitem{jones02}
Kang, H., Jones, T.W., and Gieseler, U.D.J., 2002, ApJ 579, 337

\bibitem{bl}
Lucek, S.G. and Bell, A.R., 2000, Astrop. \& Space Sc. 272, 255

\bibitem{bl1}
Lucek, S.G. and Bell, A.R., 2000a, MNRAS 314, 65

\bibitem{bl2}
Bell, A.R., and Lucek, S.G., 2001, MNRAS 321, 433.

\bibitem{malkov1}
Malkov, M.A., 1997, ApJ 485, 638

\bibitem{malkovinj}
Malkov, M.A., 1998, Phys. Rev. E 58, 4911

\bibitem{malkov2}
Malkov, M.A., Diamond P.H., and V\"{o}lk, H.J., 2000, ApJ Lett. 533, 171 

\bibitem{md01}
Malkov, M.A., and Drury, L.O'C., 2001, Rep. Prog. Phys. 64, 429

\bibitem{mal_vol}
Malkov, M. A., and V\"{o}lk, H. J., 1995,  A\&A 300, 605

\bibitem{mal_vol1}
Malkov, M. A., and V\"{o}lk, H. J., 1998, Adv. Sp. Res 21, 551

\bibitem{mond}
Mond, M. and O'C. Drury, L., 1998, A\&A 332, 385

\bibitem{ptuskin1}
Ptuskin, V.S. and Zirakashvili, V.N., 2003, A\&A 403, 1

\bibitem{ptuskin2}
Ptuskin, V.S. and Zirakashvili, V.N., 2003a, preprint astro-ph/0306226

\bibitem{topt}
Toptygin, I. N., 1999, Astronomy Lett. 25, 34

\end{thebibliography}
\end{document}